\documentclass{aa}  

\usepackage{graphicx}
\usepackage{txfonts}
\usepackage[colorlinks=true,urlcolor=blue,linkcolor=blue,citecolor=blue]{hyperref}
\usepackage[inline]{enumitem}
\usepackage{caption}
\usepackage{subcaption}
\usepackage{amsmath}

\begin{document}

    \title{Stellar mass and morphology segregation in pairs and multiplets in the cosmic web}

%    \subtitle{}

   \author{G. Torres-Ríos\inst{1}, 
           S. Verley\inst{1, 2} \and
           I. Pérez\inst{1, 2} \and
           M. Argudo-Fernández\inst{1,2} \and
           B. Bidaran\inst{1} \and
           S. Duarte Puertas\inst{1, 2} \and \\
           Y. K. González-Koda\inst{1}
          }

   \institute{Departamento de Física Teórica y del Cosmos, 
      Universidad de Granada, 18071 Granada, Spain\\
      \email{gloriatr@ugr.es}
      \and
      Instituto Carlos I de F\'isica Te\'orica y Computacional, Facultad de Ciencias, 18071 Granada, Spain
             }

   \date{Received March 17, 2026; accepted April 29, 2026}

  \abstract
  % context heading (optional)
  {} %leave it empty if necessary
  % aims heading (mandatory)
   {In this work, we investigate if the location of galaxies within the large-scale structures (LSS) of the Universe leaves an imprint in their stellar mass ($M_\star$) and morphology. To this end, we attempt to disentangle the effects of local and large-scale environments in their distributions.}
  % methods heading (mandatory)
   {We classify a sample of 25\,309 galaxies in the redshift range ${0.02 < z \leq 0.04}$ with $\log M_\star/\rm{M}_\odot \geq 9.5$ in terms of both the main LSS environments (voids, clusters, and not clusters nor voids, referred as NCNV) and local environment (singlets and multiplets; galaxies with and without companions). We present the stellar mass and morphology distributions in these environments, as well as for a subsample of galaxy pairs.}
  % results heading (mandatory)
  {Even in voids, we find that $\sim22\%$ of galaxies have companions. Stellar mass distributions show that galaxies are significantly less massive in voids, regardless of their local environment. Satellites in voids are, too, less massive with respect to their centrals than in pairs of NCNV. In terms of morphology, the denser the LSS, the greater is the proportion of early-type galaxies, even among singlets. In voids and NCNV, late-type multiplets tend to be later-type spirals than singlets. In pairs, centrals tend to be more early-type than satellites. The sample, curated to avoid morphology incompleteness, results in slightly higher fractions of early-type galaxies and multiplets than previous studies.}
  % conclusions heading (optional), leave it empty if necessary
   {We conclude that local environment alone is not sufficient to explain the distribution of stellar mass and morphology of galaxies in the local Universe. The observed mass distributions are consistent with a scenario in which the assembly of galaxies depends critically on their host halos, and the properties of these halos are related to their large-scale environment. This would explain the finding of lower-mass galaxies in voids than in denser environments, and provide a basis for considering a common evolutionary origin for multiplets.}


   \maketitle

\section{Introduction}

Looking far away enough, one can see that galaxies do not distribute homogeneously in the Universe; rather, they form a hierarchical, web-like structure along immense expanses of space. This is what we call the large-scale structure of the Universe (LSS). Within the $\Lambda$ Cold Dark Matter ($\Lambda$CDM) paradigm, the structure formation that leads to the creation of the LSS is driven by the inherent anisotropy of gravitational collapse \citep{1970A&A.....5...84Z, 1980lssu.book.....P, 1996Natur.380..603B}. The development of the LSS is therefore tied to the collapse of dark matter overdensities into halos, that would then provide enough gravitational support for cool gas to fall in and form stars -- then galaxies. Works on simulations reveal that dark matter halos residing different LSS environments possess distinct properties, such as a different halo occupation distribution, and younger halos in the most underdense environments \citep{2007MNRAS.375..489H, 2017MNRAS.470.3720T, 2018ApJ...853...84Z, 2018MNRAS.480.3978A, 2020A&A...638A..60A}. It is therefore conceivable that the distinct properties of galaxies in the LSS can be partially explained by the nature of their host dark matter halos. In other words, the location of galaxies in the Universe might be a determining factor in their properties.

Many studies report differences in galaxies inhabiting different large-scale environments. When depicting the LSS, the community typically differentiates between voids, filaments and walls, and clusters or nodes. Only around a $10\%$ of the existing galaxy mass resides in voids, despite them conforming the $\sim$70\% of the volume of the Universe \citep{2012MNRAS.421..926P, 2014MNRAS.441.2923C, 2018MNRAS.473.1195L}; and the galaxies inhabiting these extreme environments are bluer, of later morphologies, and richer in gas than galaxies in denser environments \citep{2004ApJ...617...50R, 2005ApJ...624..571R, 2007ApJ...658..898P, 2012AJ....144...16K, 2021ApJ...906...97F, 2023MNRAS.521..916R, 2024A&A...692A.258A}. The galaxy population inhabiting clusters is nothing alike; they are on average redder, of earlier morphologies, and less star-forming than galaxies in the field \citep{1980ApJ...236..351D, 1984ApJ...285..426B, 1999ApJ...527...54B, 2003MNRAS.346..601G}. Most of the galaxies in the Universe reside in intermediate densities: denser than the pristine, hollow voids, yet less dense than the hot, intracluster environment. These galaxies are usually referred to as field galaxies, but could be associated, to some extent, to filaments and walls in the cosmic web. Filaments and walls have been, too, object of study to constrict the effect of environment on galaxy evolution, finding that galaxies tend to be redder and more massive towards the spines of the large-scale filaments and nodes, where local densities are higher \citep{2014MNRAS.438.3465T, 2016MNRAS.455..127M, 2017MNRAS.466.1880C, 2017A&A...600L...6K, 2018MNRAS.474..547K, 2020A&A...639A..71K, 2022A&A...657A...9C, 2024MNRAS.529.2595B, 2024MNRAS.534.1682O}.

Nevertheless, it is not straightforward to associate this variability with the host LSS since, at local scales, the environment is also known to shape galaxy evolution. In clusters, the vast potential well, high velocity dispersions, and hot intracluster environment are the root of many local processes that are responsible of quenching and shaping the morphology of galaxies \citep{1980ApJ...236..351D}, such as harassment or ram pressure stripping \citep{1996Natur.379..613M, 2015A&A...576A.103B}. In filaments and walls, the abundance of galaxies also creates optimal conditions for galaxy tidal interactions or accretions. In cosmic voids, however, galaxies are typically more isolated than galaxies in filaments, walls, or clusters, making local environment a less likely evolutionary mechanism. This fact has made them of particular interest in the study of secular evolution. Nonetheless, it is not clear whether their different properties are due to their relative isolation, or other large-scale factors are at play. Galaxy pairs are key to comprehend the reach of the local environment in galaxy evolution, as they represent the simplest scenario in this regard. Many studies have focused on studying the visible effects of the pair environment in galaxies, including the extent of induced or suppression of star formation in pairs at different separations and stellar mass ratios \citep{2008AJ....135.1877E, 2011MNRAS.412..591P, 2012MNRAS.426..549S, 2013MNRAS.433L..59P, 2015MNRAS.452..616D, 2019MNRAS.483.5444D}. Evidence of induced star formation was identified in pairs up to separations of $\sim150 \, \mathrm{kpc}$ \citep{2013MNRAS.433L..59P}, thereby demonstrating that this alteration can occur in the absence of disruptive forms of interaction. This finding implies that the absence of a merger cannot guarantee that local factors are not in play in the evolution of a galaxy system. This is particularly relevant when trying to disentangle the impact of local and large-scale factors, since understanding the effects of local densities is the key to isolating large-scale environmental effects. 

In this work, we investigate the impact of large-scale environments on the distribution of mass and morphology, as opposed to the effects of local scales. The aim is to take a further step towards understanding how much a galaxy's characteristics are determined by its location within the cosmic web, thereby isolating the impact of local environmental factors. The galaxies in the sample are classified in terms of environments on both large and small scales, and their distribution of stellar masses and morphological types are characterised. Furthermore, in order to explore satellites in different LSS environments, a subsample of pairs in voids and in intermediate large-scale densities are presented and studied. The present study considers galaxies in multiplets as those with companions, many of those at large separations by the standards of previous studies on galaxy pairs and groups. These galaxies, while potentially bound in physical terms, do not necessarily exhibit interaction patterns. Under this definition, galaxies lacking companions are therefore considered to be unaffected by local environmental influences. This work is complementary to \cite{2024A&A...691A.341T}, where we inspected star formation histories for a sample with the same environmental classification.
    
This study is organised as follows. In Sect. \ref{sec:datasample}, we present the data and methods to characterise local and large scale environment, as well as the final samples. The results and consequent discussions are presented in Sect. \ref{sec:results} and \ref{sec:discussion}, respectively. Lastly, a summary and the main conclusions of the study can be found in Sect. \ref{sec:conclusions}. Throughout this work, a cosmology with ${\Omega_\Lambda = 0.7}$, ${\Omega_m = 0.3}$ and ${H_0 = 67.8 \text{ km} \text{ s}^{-1} \text{ Mpc}^{-1}}$ is assumed.

\section{Data and sample}\label{sec:datasample}

The employed data is presented in Sect. \ref{sec:data}, while the selection of the sample is detailed in Sect. \ref{sec:environment} and \ref{sec:completeness}. In the prior we detail the galaxy environment characterisation, both in the large and the local scales, and in the latter we discuss the completeness of the sample.

 \subsection{Data}\label{sec:data}
    
    This work makes use of the galaxy group catalogues released in \citet{2017A&A...602A.100T}, which are based on the Sloan Digital Sky Survey Data Release 12 \citep[SDSS DR12;][]{2015ApJS..219...12A} data. Out of these, the galaxy catalogue contains 584\,449 galaxies in $0.0<z<0.2$, and provides positions, redshifts and other spectroscopic features, as well as group membership. This catalogue provides a clean, large sample appropriate for galaxy environment research. Only galaxies with Petrosian $r<17.77\:\mathrm{mag}$ were selected, according to the SDSS completeness limit \citep{2002AJ....124.1810S}. All redshifts used in this work are corrected by cosmic microwave background effect ($z_\mathrm{CMB}$). Further details on the refinement of the initial sample can be found in Section 2 of \cite{2017A&A...602A.100T}.

    All stellar masses ($M_\star$) were obtained from the MPA-JHU spectroscopic catalogue \citep{2003MNRAS.341...33K, 2007ApJS..173..267S}, through the CasJobs tool\footnote{\url{https://skyserver.sdss.org/casjobs/}}.

    To explore the morphology of the galaxies in the sample, we use the catalogue provided in \cite{2018MNRAS.476.3661D}, which is based on the automated classification of galaxies using deep learning on \textit{gri} images from the SDSS DR7 \citep{2009ApJS..182..543A}. We use T-Type as morphology indicator, which is a continuous variable directly associated with the classical Hubble sequence classification. A $\text{T-Type}\leq0$ corresponds to early-type galaxies (comprising elliptical and S0 galaxies), and $\text{T-Type}>0$ are spiral galaxies (from Sa to Sm). Inspecting the probability of being an S0 galaxy ($P_{\mathrm{S0}}$) as provided, too, by \cite{2018MNRAS.476.3661D}, we find that $79\%$ of the early-type galaxies in our initial sample have S0 morphologies ($P_{\mathrm{S0}}>0.5$). However, these galaxies are not considered separately, as we discuss early-type galaxies as a whole throughout this work. The galaxies in our sample have T-Type values ranging from $-2.8$ to $6.9$, as will be shown in detail in Section \ref{sec:results-local}.
    
    Out of the total of 584\,449 galaxies, we restrict the sample to galaxies of masses $\log M_\star/\rm{M}_\odot > 9.5$ within the redshift range $0.02<z\leq0.04$ (the choice of these limits is discussed in Section \ref{sec:completeness} and Appendix \ref{appendix:et-bias}). This reduces the sample to 25\,309 galaxies pendant of both local and LSS environment classification.

\subsection{Classification of environment}\label{sec:environment}

\begin{figure*}
    \centering
    \includegraphics[width=0.495\linewidth]{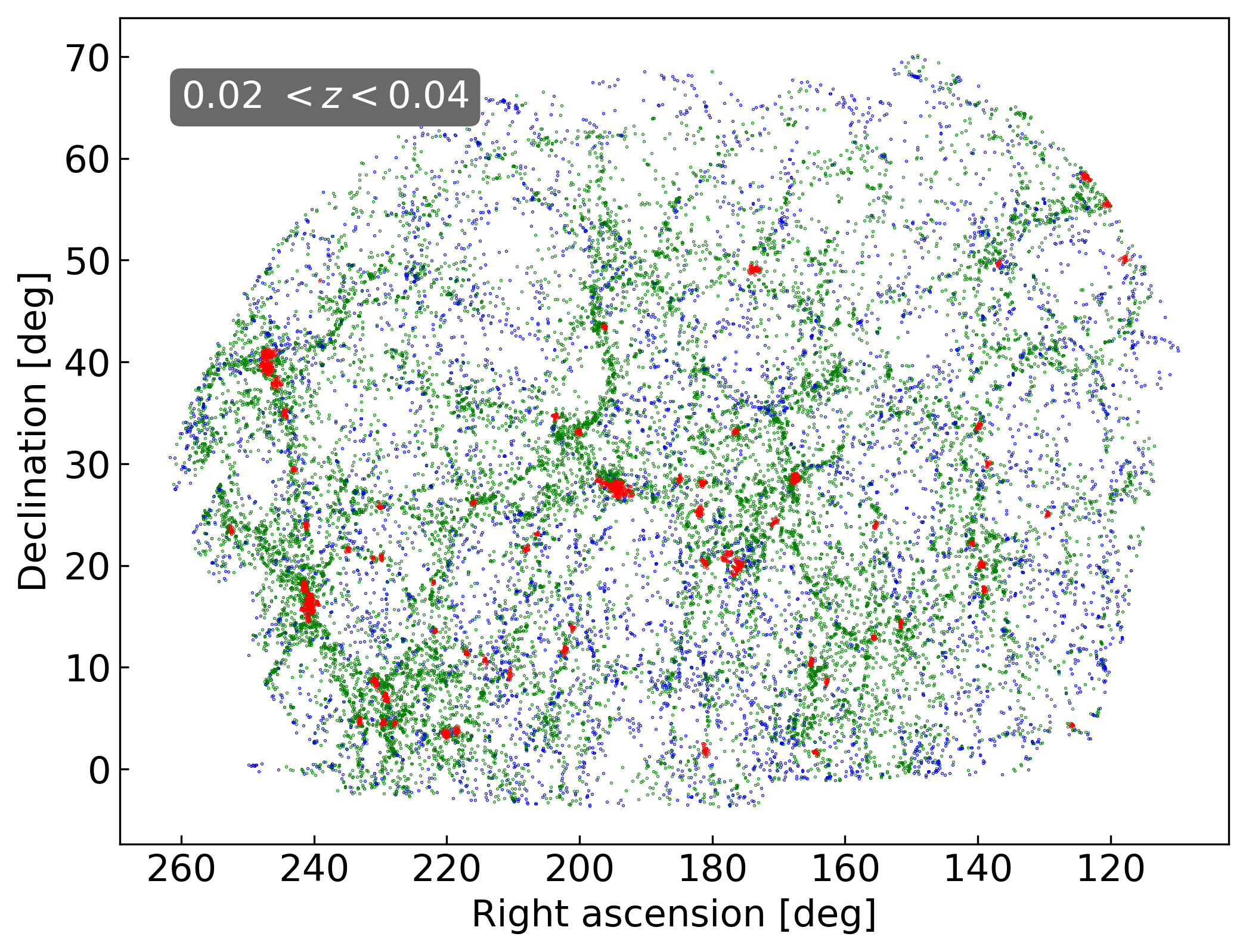}
    \includegraphics[width=0.45\linewidth]{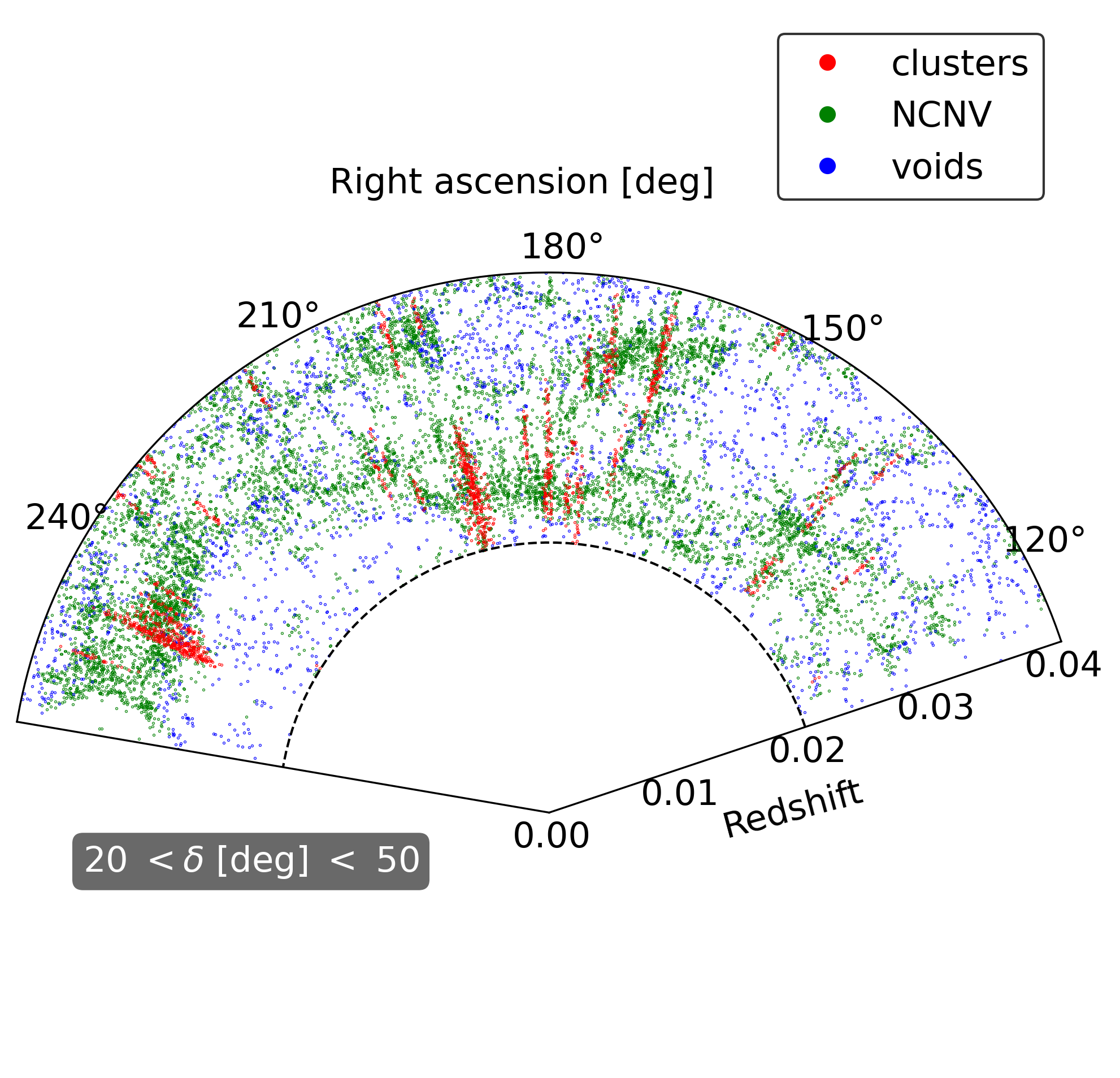}
    \caption{Spatial distribution of galaxies in the sample. \textit{In the left panel}, a slice in redshift ($0.02 < z < 0.04$, comprising the whole sample) in the right ascension-declination space. \textit{In the right panel,} a slice in declination ($20 < \delta \: \mathrm{[deg]} < 50$) in the right ascension-redshift space. The colour-coding is associated to the LSS classification of the sample (voids in blue, NCNV in green, and clusters in red). The most local galaxies with $z<0.02$ (under the black dashed line in the right panel) were excluded from this study, as their LSS classification is compromised. This representation mimics the ones provided by LSSGalPy \citep{2017PASP..129e8005A}, a 3D LSS visualisation tool publicly available at \url{https://gitlab.com/astrogal}.}
    \label{fig:sample-example}
\end{figure*}

Firstly, we categorise galaxies according to their position within the large-scale structures of the Universe. We follow the methodology previously applied by \cite{2023Natur.619..269D} and \cite{2024A&A...691A.341T}, which we elaborate below. 

On the one hand, galaxies in voids are identified through crossmatching with the void galaxy catalogue provided by \cite{2012MNRAS.421..926P}. This catalogue is based on the application of the VoidFinder algorithm \citep{2002ApJ...566..641H} on SDSS DR7 data, and is successful in identifying void galaxies up to $z=0.10$. However, at very low redshifts the effect of peculiar velocities and large uncertainties in photometric data may affect also the LSS classification. At $z<0.02$, there is a significant drop in the number of objects and an atypical fraction of galaxies are identified as void galaxies, most of which exhibit some sort of association to structure when visually checked. In light of the aforementioned points, we choose to exclude galaxies with $z<0.02$. On the other hand, galaxies in clusters are identified as those being part of groups of 30 galaxies or more \citep{1989ApJS...70....1A}, according to the classification in groups performed by \cite{2017A&A...602A.100T}. The rest of the galaxies will be referred to as NCNV (not in voids nor clusters) galaxies henceforth, and were denoted as filament galaxies in \cite{2024A&A...691A.341T}. A small subset of galaxies (161; a 0.6\% of the sample) is both in a void and a cluster according to our previous criteria, and are discarded of this work. This classification in LSS is portrayed in Figure \ref{fig:sample-example}, where we show the spatial distribution of galaxies in the sample. As can be seen in the Figure, many NCNV galaxies form part of filamentary structures. We do not characterise filaments and walls as cosmic web structures, although it is noteworthy that NCNV galaxies somewhat represent the range of intermediate large-scale density that is associated with filaments and walls. The full sample of 25\,309 galaxies within ${0.02<z\leq0.04}$ is divided into 2\,787 cluster galaxies \citep[11.0\%, distributed in 89 clusters, according to][]{2017A&A...602A.100T}, 16\,984 NCNV galaxies (67.5\%), and 5\,377 void galaxies \citep[21.4\%, distributed in 97 distinct voids according to][]{2012MNRAS.421..926P}. Only 161 galaxies (0.6\% of the sample) are removed due to conflicted characterisation, as discussed earlier, leaving a usable sample of 25\,148 galaxies.

For the classification in local environment, and independently of the LSS characterisation, companion galaxies were searched for every galaxy within the whole galaxy catalogue provided by \cite{2017A&A...602A.100T}. With that purpose, we consider galaxy companions those within velocity differences $\Delta v \leq 160 \ \mathrm{km \ s}^{-1}$ and sky-projected distances $\Delta r_p \leq 0.45 \ \mathrm{Mpc}$ from a particular galaxy. In \cite{2015A&A...578A.110A}, these conditions were found to assure physical bounding in the study of isolated pairs and triplets. In non-isolated environments, these conditions may extend to larger $\Delta v$ and $\Delta r_p$ and therefore, they delimit a conservative approach for considering galaxies with companions in this study. As these conditions do not imply isolation or the absence of it, we call galaxies devoid of companions in the $\Delta v - \Delta r_p$ space "singlets", and galaxies with companions are called "multiplets" \citep[these same definitions were introduced in ][where multiplets were referred as "group members"]{2024A&A...691A.341T}. Take into consideration that multiplets are not necessarily part of physically bound groups, and this definition simply denotes those galaxies which have companions within the specified $\Delta v$ and $\Delta r_p$ limits. The main sample (i.e., the 25\,148 galaxies properly characterised in LSS) is made of 12\,393 multiplets (49.3\%), and 12\,755 singlets (50.7\%). 

For further analysis, we create a subsample of galaxies in pairs, which are included in the sample of multiplets by construction. These galaxies have a single neighbour, and this companion is not physically bound to any other galaxy. To ensure the cleanliness of the analysis, we only consider pairs of galaxies that are both classified as inhabiting the same LSS. In all cases, the galaxy in the pair with the highest stellar mass is defined as the central. The pair sample contains 1\,108 pairs, with 1\,006 pairs having both galaxies inhabiting the same, well-characterised, LSS environment. Out of those, 199, 739, and 68 pairs inhabit voids, NCNV, and clusters, respectively. 

The sample of pairs in clusters, however, will not be discussed in this paper. Although we classify local and LSS environments separately, clusters are extreme environments with a large gravitational potential that bounds together most of the galaxies inhabiting them, making the local- and large-scale phenomena hard to differentiate. Our notion of a galaxy in a pair is, essentially, a galaxy with a single physically bounded companion. This idea is not compatible with the pair inhabiting an environment in which the physical boundary goes beyond the galaxy scales, by definition. For this reason, we will focus on the comparison of void and NCNV galaxies when discussing galaxies in pairs.

\subsection{Completeness}\label{sec:completeness}

\begin{figure}
    \centering
    \begin{subfigure}[t]{0.49\textwidth}
    \includegraphics[width=1\linewidth]{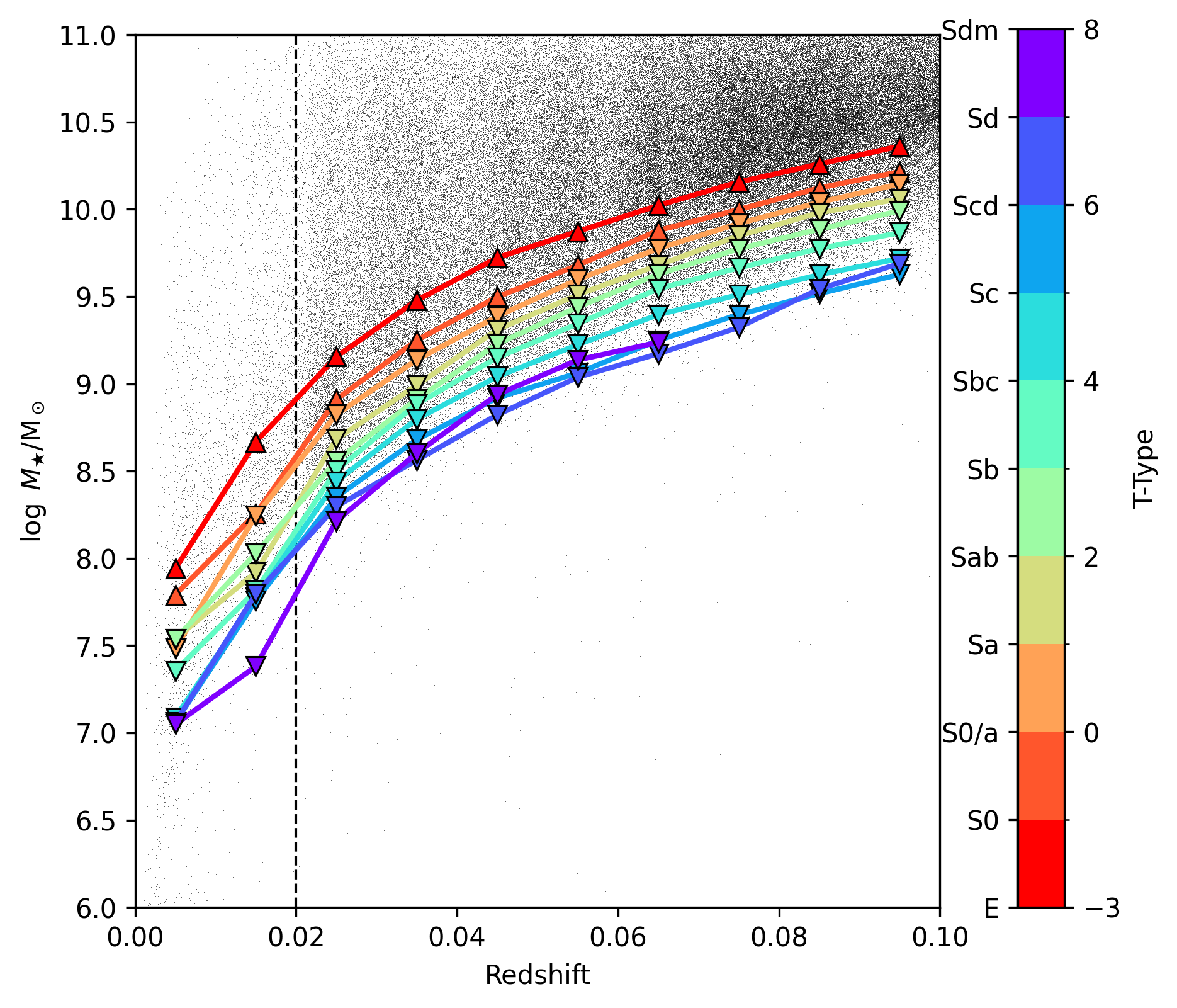}
    \caption{Data binned and colour-coded by T-Type (morphology), provided by \cite{2018MNRAS.476.3661D}. Triangle markers denote the curves of early-type galaxies, while inverted triangles denote late-types.}    
    \label{fig:completeness_TType}
    \end{subfigure}
    \begin{subfigure}[t]{0.49\textwidth}
    \includegraphics[width=\linewidth]{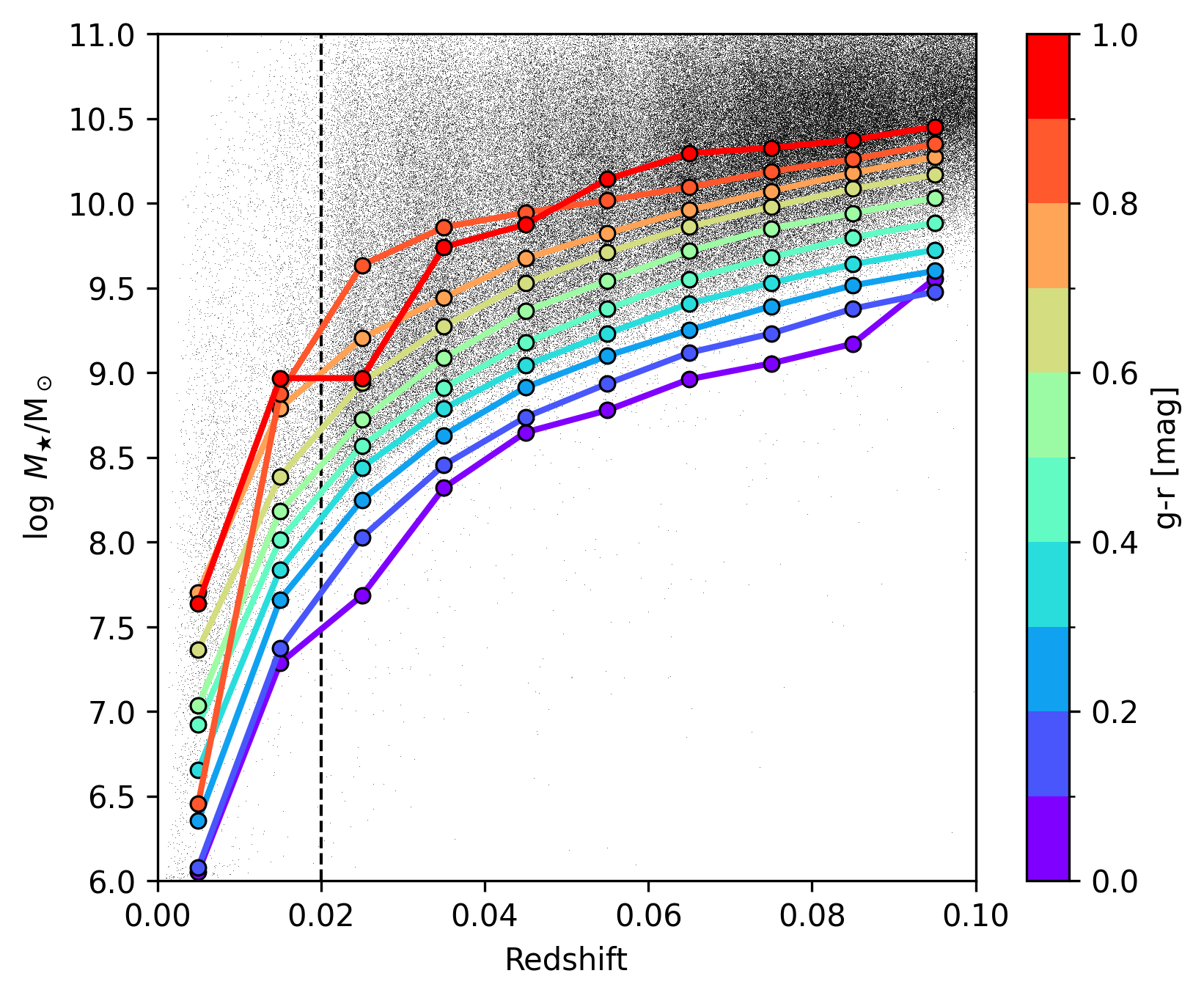}
    \caption{Data binned and colour-coded by $g-r$ colour, provided by the SDSS DR12.}
    \label{fig:completeness_color}
    \end{subfigure}
\caption{Stellar mass as a function of redshift for the full parent sample. The shown curves portray completeness for different bins in T-Type (panel \ref{fig:completeness_TType}) and $g-r$ colour (panel \ref{fig:completeness_color}), as shown by the colourbars. The completeness is computed as the 5\textsuperscript{th} percentile of the stellar mass in bins of redshift of width 0.01. Dashed vertical lines mark our lower limit in redshift, set by the reliability of the LSS characterisation.}
\label{fig:completeness}
\end{figure}

The initial sample (galaxies in \cite{2017A&A...602A.100T} catalogue within $z<0.1$) is shown in Figure \ref{fig:completeness}, in the $z$--$\log M_\star$ space. Completeness, computed as the 5\textsuperscript{th} percentile of stellar mass (corresponding to the 95\% of completeness) per redshift bin, is represented in terms of T-Type (upper panel) and $g-r$ colour (lower panel), both parameters being key galaxy properties. As discussed in the previous section, the drop in object number and possible misclassification of void galaxies in $z<0.02$ is clearly visible on the left side of the dashed vertical lines. Note that in Figure \ref{fig:completeness_color}, completeness within $g-r > 0.9\,\mathrm{mag}$ drops to lower masses than expected by the seen trend. This is evidence of colour not being an infallible morphology tracer: when visually inspected, we find that many of these galaxies are bright discs that have been significantly reddened by dust. This effect is not there in Figure \ref{fig:completeness_TType}, as T-Type is more descriptive of morphology than colour. Nevertheless, $g-r$ proves to be sufficient to characterise morphological completeness. 

From Figure \ref{fig:completeness}, it is straightforward to notice that completeness is contingent on morphology; this is caused by the fact that, for a given mass, red galaxies are fainter compared to blue galaxies. Moreover, red galaxies are more easily missed by the SDSS filters. It is therefore necessary to take this into account to have a complete sample in both stellar mass and morphology. We explore this further in Appendix \ref{appendix:et-bias}, and provide a parametrisation of the dependence of stellar mass completeness with morphology and redshift for this sample, but applicable to studies based on the SDSS DR12 main galaxy sample. As discussed in said Appendix, to assure completeness, in this work we make use of the sample within ${0.02 < z \leq 0.04}$ and ${\log M_\star/\mathrm{M_\odot}>9.5}$. To obtain these limits we used the T-Type-based completeness parametrisation.

\section{Results}\label{sec:results}

The results of this work are divided in three parts: in Sect. \ref{sec:results-fractions}, we provide the fraction of galaxies in the sample in each LSS and their local environments, as well as a characterisation of the selection bias in the sample. Sect. \ref{sec:results-local} explores stellar mass and morphology distributions of the sample in terms of environment. Finally, Sect. \ref{sec:results-pairs} explores the sample of pairs in each LSS, considering central and satellite galaxies separately. 

\subsection{Fractions of singlets and group members in the LSS}\label{sec:results-fractions}

\begin{table}[]
\caption{Fraction of galaxies belonging to different local environment classifications across the LSS}
    \centering
\begin{tabular}{lccccc}
\hline\hline
& & & In & In & Larger \\
LSS & Singlets & Multiplets & pairs & triplets & multiplets\\
& [\%] & [\%] & [\%] & [\%] & [\%] \\
\hline
Void & 78.1 & 21.9 & 8.3 & 0.9 & 12.7 \\
NCNV & 59.6 & 40.4 & 9.5 & 3.7 & 27.2 \\
Cluster & 22.0 & 78.0 & 5.3 & 2.2 & 70.5 \\
\hline
All & 59.4 & 40.6 & 8.8 & 2.9 & 28.9 \\
\hline
\end{tabular}
    \tablefoot{Using the complete sample of 25\,148 galaxies in the local Universe. The percentages are shown for the different LSS (void, NCNV, and cluster galaxies) as well as for the whole sample. By definition, the number of group members is the sum of galaxies in pairs, in triplets and in larger multiplets, where larger multiplets denote galaxies with more than two companions.}
    \label{tab:percentages}
\end{table}

First, we inspect the percentage of galaxies belonging to each environment, both on a large scale and a local scale basis (Table \ref{tab:percentages}). One should take into consideration that restricting the sample to high-mass galaxies converts all galaxies with low-mass companions to singlets, by construction. In this work, that definition reduces the fraction of multiplets: if companions with stellar masses $9.0 <\log M_\star/\mathrm{M}_\odot < 9.5$ were taken into consideration, the fraction of multiplets would increase $\sim$8\% in all LSS environments. As it is, the fractions of multiplets in Tab. \ref{tab:percentages} may be taken as lower limits, or simply considered as galaxies with high-mass companions.

Additionally, as the present parent sample is based on SDSS data, it should be considered that fibre collision affects the local environment classification. The minimum separation of the spectroscopic fibres is $55''$: this means that objects closer than this projected distance cannot be detected simultaneously \citep{2002AJ....124.1810S}. This limitation naturally leads to an overestimation of the number of singlets and, in the same way, what is sometimes detected as a pair may truly be a triplet containing an undetected close pair. We apply the same method as \cite{2012A&A...540A.106T} to estimate the loss of pairs due to fibre collisions. Only a 4\% of singlets have a missing neighbour in the SDSS, since most of the missing targets are multiplets. Around the 60\% of these companions are also close in redshift \citep[within $500\,\mathrm{km \cdot s^{-1}}$; ][]{2002ApJ...571..172Z} to the galaxy. Therefore, 2.4\% of singlets have missing neighbours. In the present work, that loss fraction is an overestimation due to our selection criterion of velocity differences below $160\:\mathrm{km \cdot s^{-1}}$. In any case, that estimate translates to a 1.4\% in the total sample. Breaking this down by the cosmic web environments, we estimate a 1.9\% loss of pairs in voids, 1.4\% in NCNV, and 0.5\% in clusters. Consequently, the real fraction of singlets would be 98.6\% of the one provided for the whole sample, and 98.2\%, 98.6\%, and 99.5\% for voids, NCNV and clusters, respectively.

The fraction of lost pairs by the fibre collision effect is significantly lower in all environments than the 8\% reported by \cite{2012A&A...540A.106T}. However, we tested that if our sample did not account for the morphology bias in the completeness of the sample (Appendix \ref{appendix:et-bias}), that percentage would be compatible with our pair loss estimate in the whole sample (9.2\%), and very similar to the one in NCNV (7.9\%). Accounting for that effect implied limiting the redshift and stellar mass range in the sample, and in consequence, we find a higher fraction of multiplets, as not as many galaxies are lost due to incompleteness. Since the fraction of singlets is lower, this method leads to a lower percentage of missing pairs, too.

\subsection{Singlets and multiplets} \label{sec:results-local}

\begin{figure*}
    \sidecaption
    \includegraphics[width=12cm]{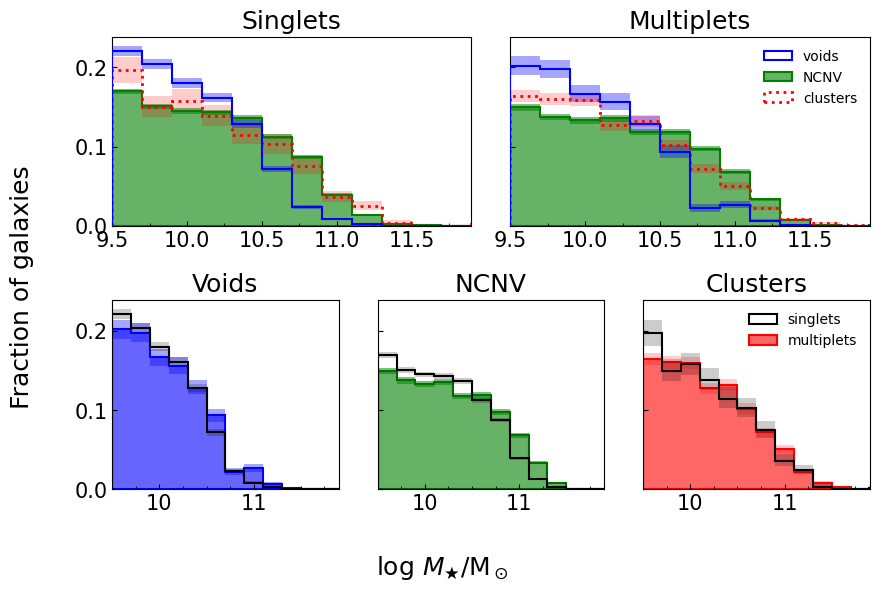}
    \caption{Stellar mass distributions for galaxies in the sample. \textit{On the upper panels}, galaxies are separated in singlets and multiplets (galaxies with companions) and colour-coded by LSS (voids in blue, NCNV in green, and clusters in red). \textit{On the lower panels}, the distributions are separated according to the LSS, and show multiplets in solid colour and singlets in black line histograms. The shaded areas correspond to $1\sigma$ bootstrapping uncertainties.}
    \label{fig:pp-mass}
\end{figure*}

\begin{figure}
    \centering
    \includegraphics[width=\linewidth]{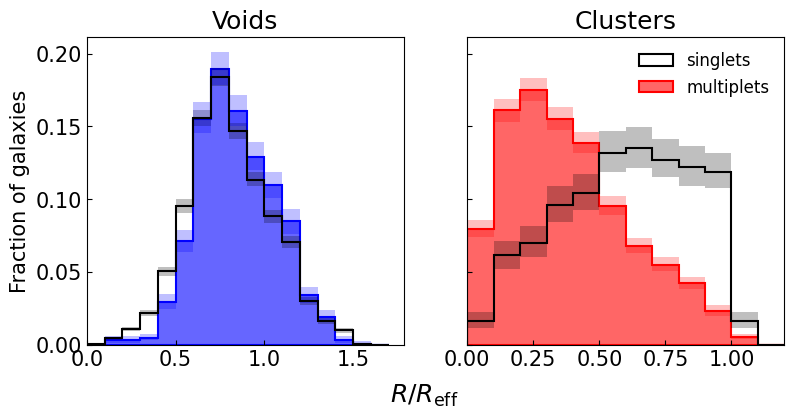}
    \caption{Distribution of sky-plane distances to the centre of the void (blue) or cluster (red) for multiplets (filled) and singlets (outlined histograms). Shaded areas portray $1\sigma$ bootstrapping uncertainties. Distance is normalised by the effective radius of the void/cluster ($R_{\rm{eff}}$). In voids, this effective radius is defined as the radius of a sphere with equal volume to the void region. Values $R/R_{\rm{eff}}>1$ are reflection of galaxies located in the outskirts of voids that are not spherical (as it is frequently the case). In clusters, the effective radius is $R_{200}$, defined as the radius in which the cluster's mean density is 200 times higher than the mean density of the Universe.}
    \label{fig:effr_frac}
\end{figure}

Figure \ref{fig:pp-mass} shows the stellar mass distributions for singlets and multiplets depending on the LSS environment. The upper panels show the distributions of stellar mass for singlets and multiplets in the three main large-scale environments. Singlets distributions have median values of $9.978 \pm 0.008$, $10.148 \pm 0.007$ and $10.09 \pm 0.02$ dex in voids, NCNV, and clusters respectively, and $10.01 \pm 0.02$, $10.218 \pm 0.009$, and $10.13 \pm 0.02$ dex in the case of multiplets, where uncertainties represent the bootstrap standard error of the median. Voids contain a higher fraction of low-mass galaxies in any local environment configuration than any other LSS environment, and NCNV galaxies show the highest fractions in high masses, as cluster galaxies (mostly multiplets, by construction) are slightly less massive on average. In the lower panels, we show that, in voids and clusters, singlets and multiplets exhibit similar distributions, with visible differences lying within the estimated uncertainties. In NCNV, singlets appear to be slightly less massive than multiplets, being the difference of the medians of the distributions 0.07. Kolmogorov-Smirnov (KS) tests reveal p-values under $0.05$ for both voids and NCNV when comparing singlets and group members distributions, but show $p=0.2$ in the case of clusters, implying that in this last case the distributions are indistinguishable. Extremely low p-values arise when comparing LSS environments with each other in either of the local environment classifications, excepting singlets in NCNV and clusters ($p=0.06$) that are, again, nearly indistinguishable. Galaxies in clusters identified as singlets are mostly in the outskirts of their corresponding cluster, as can be deduced from Figure \ref{fig:effr_frac}, where the distance distribution of the cluster singlets is biased towards the periphery. Given that the local densities in the outskirts of the clusters are close to those of filaments and walls, it is expected to find no measurable differences between cluster and NCNV singlets. 

Overall, we find that galaxies in voids tend to be, typically, less massive than galaxies in denser environments, independently of local environment. The stellar mass distribution of NCNV singlets is very similar to what is found in clusters, but multiplets tend to slightly higher masses. Attending the shape of the distributions, the difference in stellar mass distributions when inspecting local environment is subtler than the observed differences in the large-scales, that are more significant. It is possible that more significant differences in the local environment distributions could be obscured by the lower mass limit in this study. For instance, in voids, we would anticipate numerous galaxies to have low-mass companions; in this sample, these galaxies would be classified as singlets, therefore minimising the differences between the mass distributions of singlets and multiplets.

\begin{figure*}
    \sidecaption
    \includegraphics[width=12cm]{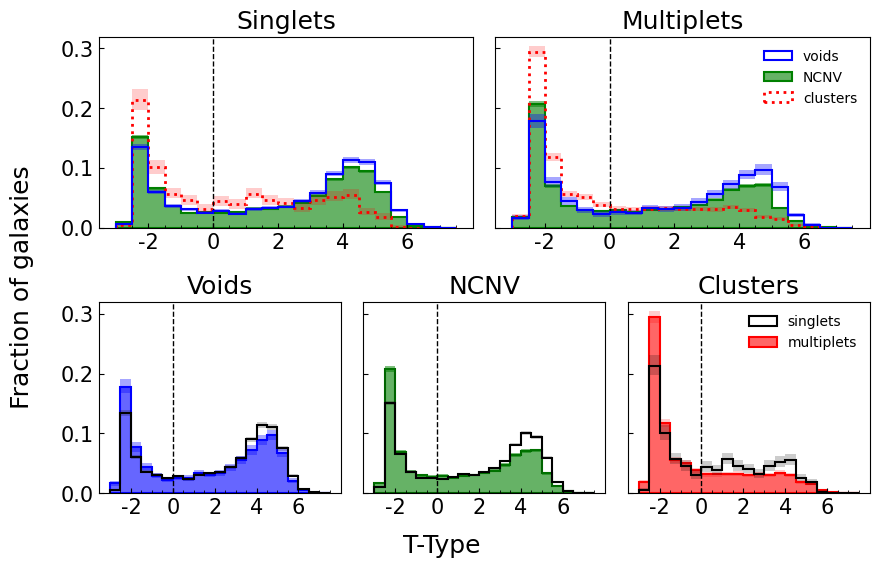}
    \caption{T-Type distributions for galaxies in the sample. \textit{On the upper panels}, galaxies are separated in singlets and multiplets (galaxies with companions) and colour-coded by LSS (voids in blue, NCNV in green, and clusters in red). \textit{On the lower panels}, the distributions are separated according to the LSS, and show multiplets in solid colour and singlets in black line histograms. The shaded areas portray $1\sigma$ bootstrapping uncertainties. The dashed vertical lines separate early- ($\text{T-Type}\leq0$) and late-type ($\text{T-Type}>0$) galaxies.}
    \label{fig:ttype-local}
\end{figure*}

Figure \ref{fig:ttype-local} shows the T-Type distributions for the singlet and multiplets samples. As described in Section \ref{sec:data}, T-Type is a continuous variable related to the Hubble sequence and is used in this work to characterise galactic morphology. On the upper panels, the distributions differ significantly across the LSS, as with environmental density comes a higher fraction of early-type galaxies. Consequently, voids exhibit the higher fractions of late-type galaxies. When comparing voids and NCNV, we find noteworthy that, whereas singlets' T-Type distributions are compatible, multiplets in NCNV show a rise in the fraction of early types. Cluster galaxies, however, exhibit a completely different T-Type distribution for any local classification, with a huge predominance of early-type galaxies. When it comes to the difference within local environments at fixed LSS, multiplets in voids are not only more early-type than singlets, but the late-types tend to have higher T-Type's, translating to more evolved spirals than in void singlets. The same occurs in NCNV environments. When interpreting these results, it must be acknowledged that the samples of different environments are not matched in stellar mass.

Ultimately, we find that there is a higher fraction of early-type galaxies being multiplets than singlets in every LSS environment. This result is better seen in Table \ref{tab:ttype-local}, that shows the percentage of multiplets (or singlets) that are early- or late-type. Considering the whole sample, multiplets are 14.1\% more inclined to be early-type than singlets. Those fractions are 7.9\%, 10.0\%, and 14.1\% when broken down into galaxies in voids, NCNV, and clusters, respectively. In clusters, it is significant that, despite the predominance of early-types, more than 35\% of multiplets are late-type. These late-type galaxies, however, appear to be equally distributed in the $\text{T-Type}>0$ range, in opposition to late-types in voids and NCNV, that peak around $\text{T-Type}=4.5$ (roughly Sc galaxies).

\begin{table}[]
\caption{Fraction of early- and late-type galaxies within the samples of singlets and multiplets across the LSS}
    \centering
\begin{tabular}{lcccc}
\hline \hline
 & \multicolumn{2}{c}{Singlets} & \multicolumn{2}{c}{Multiplets} \\
LSS & Early-type & Late-type & Early-type & Late-type \\
 & [\%] & [\%] & [\%] & [\%] \\
\hline
Void & 30.1 & 69.9 & 38.0 & 62.0 \\
NCNV & 34.2 & 65.8 & 44.3 & 55.7 \\
Cluster & 49.7 & 50.3 & 64.1 & 35.9 \\
\hline
All & 33.6 & 66.4 & 47.8 & 52.2 \\
\hline
\end{tabular}
    \tablefoot{Early-type galaxies are those with $\text{T-Type}\leq0$, and late-type galaxies those with $\text{T-Type}>0$. The percentages are shown for the different LSS (void, NCNV, and cluster galaxies) as well as for the entire sample, and are given with respect the amount of galaxies in a fixed LSS and local environment configuration (e.g., adding the percentages of early- and late-types in void singlets results in 100\%).}
    \label{tab:ttype-local}
\end{table}

\subsection{Galaxy pairs}\label{sec:results-pairs}

The pair samples contain 199 and 739 pairs in voids and NCNV, respectively. As noted in Section \ref{sec:environment}, in this work, pairs are selected by finding neighbours within velocity differences $\Delta v \leq 160 \ \mathrm{km \ s}^{-1}$ and sky-projected distances $\Delta r_p \leq 0.45 \ \mathrm{Mpc}$, then restricting the number of neighbour to consider pairs. For the sake of analysis, we restrict the sample to pairs with the same LSS classification. In Appendix \ref{appendix:pairs}, we show some examples of pairs (Figure \ref{fig:examples-pairs}) and their distributions of $\Delta v$ and $\Delta r_p$ (Figure \ref{fig:tempel-pp-dv-pd}).

\begin{figure}
    \centering
    \includegraphics[width=\linewidth]{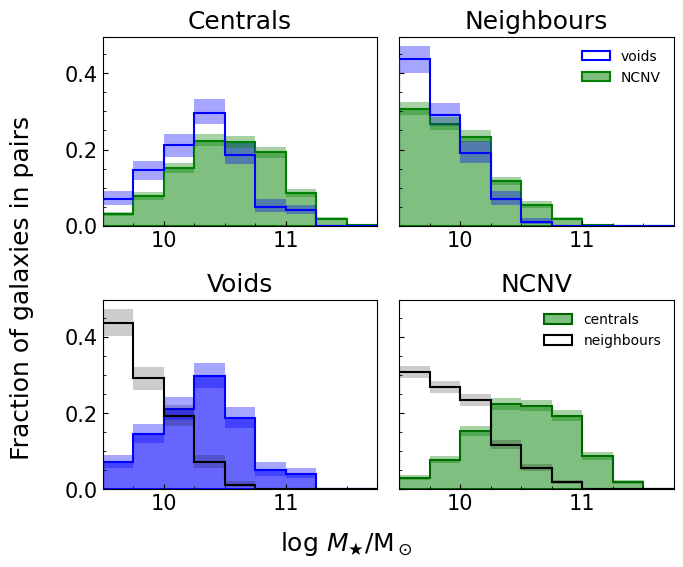}
    \caption{Stellar mass distributions for galaxies in pairs. \textit{On the upper panels,} separated in central and neighbour distributions and colour-coded by LSS (voids in blue and NCNV in green). \textit{On the lower panels}, distributions of centrals (solid colour) and neighbours (black lines) for pairs in voids and NCNV. The lower limit of $10^{9.5}$ M$_\odot$ was manually set to assure the completeness of the sample. The presented samples contain 199 and 739 pairs in voids and NCNV, respectively. Shaded areas show $1\sigma$ bootstrapping uncertainties.}
    \label{fig:m_cn_distrib}
\end{figure}

\begin{figure*}
    \sidecaption
    \includegraphics[width=12cm]{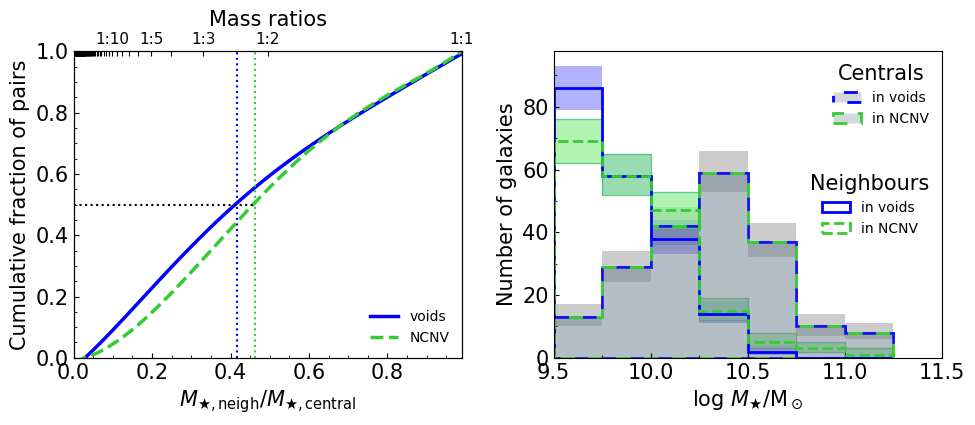}
    \caption{Comparisons of the mass of neighbours ($M_{\rm{neigh}}$) to the mass of the centrals ($M_{\rm{central}}$) in pairs in voids (blue) and NCNV (green). \textit{On the left panel}, cumulative distribution of the neighbour-to-central stellar mass fraction ($M_{\rm{neigh}}/M_{\rm{central}}$). The medians of the distributions are marked with dotted lines. \textit{On the right panel,} distributions of stellar mass in central and neighbour galaxies. The shaded areas portray $1\sigma$ bootstrapping uncertainties. These subsamples in both large-scale environments have, by construction, the same number of galaxies (198) and the same stellar mass distribution for the central galaxies (in grey, on the right-hand panel).}
    \label{fig:mn_mc}
\end{figure*}

Figure \ref{fig:m_cn_distrib} shows the stellar mass distributions of central and neighbour galaxies in voids and NCNV, where the central galaxy is defined as the most massive in the pair. Considering the median values of the $\log M_\star/\mathrm{M}_\odot$ distributions, voids have a median stellar mass of $10.32\pm0.03$ and $9.78\pm0.02$ dex in centrals and neighbours respectively; while in NCNV $10.52\pm0.02$ and $9.93\pm0.02$ dex, where uncertainties denote the bootstrap standard error of the median. The difference between medians is very similar in both voids and NCNV (0.54 and 0.59 dex). It must be taken into consideration, though, that the completeness lower limit of $10^{9.5}\,\mathrm{M_\odot}$ heavily constricts the dynamical range of the mass distributions. It is clear, nonetheless, that the centrals and neighbours stellar mass distributions vary in different LSS ($p<10^{-5}$ when running a KS test): in void pairs, both centrals and neighbours are less massive than the ones inhabiting NCNV.

Aiming to study the relation between central and neighbour stellar masses further, we create subsamples of the two LSS environments that have the same stellar mass distributions of the central galaxies. To do this, we compute a pairwise crossmatch. For each void central, we find the central in NCNV that has the closest stellar mass. If the difference between masses is within 0.1 dex, we keep both pairs. We consider the void pairs sample as a reference, as it is the smallest of the two samples. Using this method, we end up with subsamples of 198 pairs for each LSS environment that have a nearly identical distribution of stellar mass for central galaxies -- KS tests reveal p-values over 0.99 when comparing these distributions, and only three of the matches have a stellar mass difference over 0.025 dex.

Using these samples, we compare the mass distributions of satellites and centrals in the LSS in Figure \ref{fig:mn_mc}. We plot the neighbour-to-central stellar mass fraction ($M_n/M_c$) on the left panel, and the distributions of stellar mass of neighbours on the right panel. Since central galaxies are defined as the most massive of the pair, $M_n/M_c$ is normalised to the $[0,1]$ range by construction. We find that pairs in void galaxies tend to have lower $M_n/M_c$ than galaxies in NCNV, i.e., larger stellar mass differences between the central and the neighbour in pairs. The median of the cumulative distributions (marked by dotted lines in left-hand panel) correspond to mass ratios of 1:2.4 and 1:2.2 in voids and NCNV, respectively. We previously found that the difference of medians is similar in voids and NCNV when comparing the stellar mass distribution of centrals and neighbours. However, the difference that we find in $M_n/M_c$ may indicate that the medians of the distributions are not representative enough. This is due to the limited dynamical range of stellar mass caused by the $M_\star > 10^{9.5} \, \mathrm{M}_\odot$ restriction in our sample.

\begin{figure}
    \centering
    \includegraphics[width=\linewidth]{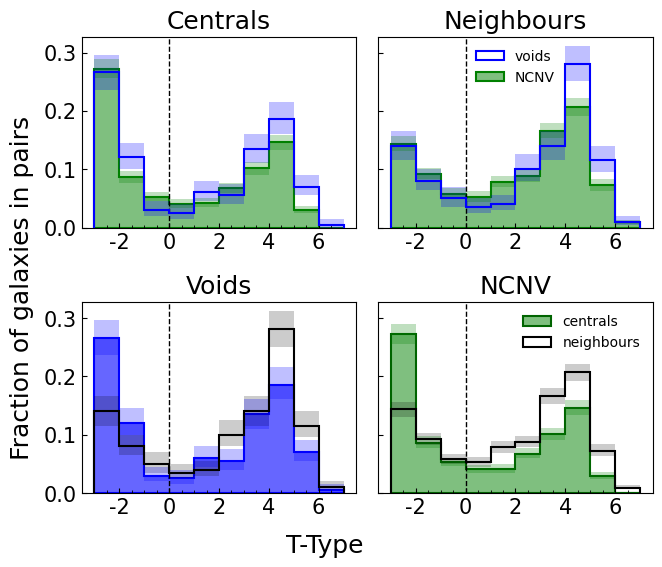}
    \caption{T-Type distributions for galaxies in pairs. \textit{On the upper panels,} the distributions are separated in central and neighbour distributions and colour-coded by LSS (voids in blue, NCNV in green). \textit{On the lower panels}, distributions of centrals (solid colour) and neighbours (black lines) for every LSS. The presented samples contain 199 and 739 pairs in voids and NCNV, respectively. Shaded areas portray $1\sigma$ bootstraping uncertainties.}
    \label{fig:ttype-pairs}
\end{figure}

\begin{figure}
\centering
\includegraphics[width=\linewidth]{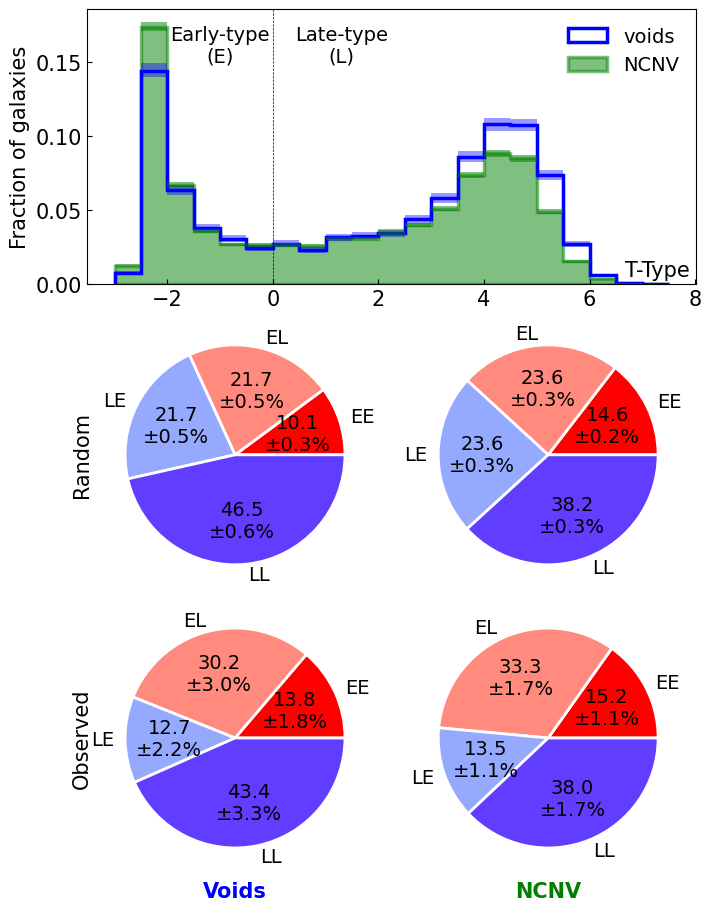}
 \caption[]{Morphological segregation in pairs in the cosmic web. \textit{From bottom to top,} \textit{the bottom row} of pie charts shows the morphological combinations in observed pairs. From left to right, charts are shown for voids and NCNV. \textit{The top row of pie charts} shows a simple random sampling model, taking into account the fraction of early- and late-type galaxies in each LSS environment. These fractions are obtained from the T-Type distributions shown in the top histograms, that consider all galaxies in the parent sample that inhabit voids (blue) and NCNV (green). Uncertainties are estimated assuming a binomial distribution for $P(\rm{E})$ and $P(\rm{L})$ and propagating quadratically for the different morphological combinations. In the upper histograms, shaded areas portray $1\sigma$ bootstrapping uncertainties.}
 \label{fig:piecharts}
\end{figure}

In order to address the study of morphology in pairs (making use of the pairs sample, not matched in stellar mass of the central galaxy), the T-Type distributions within pairs are shown in Figure \ref{fig:ttype-pairs}. When comparing cosmic web environments, galaxies in pairs replicate what was laid down in Figure \ref{fig:ttype-local}: galaxies in voids tend to have a later T-Type, both in centrals and satellites, than galaxies in NCNV. The difference between centrals and satellites is, too, significant. In both cosmic web environments, central galaxies tend to express earlier-type morphologies (lower T-Type) when compared to neighbour galaxies. In voids, 56.3\% of pairs have late-type central galaxies, and 72.7\% of late-type neighbours. In NCNV, too, most of the centrals and neighbours are late-type (51.0\% and 69.7\%, respectively). Comparing the T-Type distributions in pairs with the distributions for multiplets in the whole sample (Figure \ref{fig:ttype-local}), we find that, although multiplets (including galaxies in pairs) are, overall, more early-type, the mass segregation when dividing pairs in centrals and neighbours is enough to tip the scale towards later types in satellites.

To study in more detail how morphology in pairs happens in the context of the LSS, we examine the combinations of early- and late-type morphologies in each environment. We classify pairs as LL, LE, EL, and EE, where L and E denote late-types ($\text{T-Type}>0$) and early-types ($\text{T-Type}\leq0$), respectively. The first letter corresponds to the central galaxy, while the second is the neighbouring companion (e.g., a pair with a late-type central and an early-type neighbour would be tagged as LE).

As a reference baseline model, we simulate the formation of a pair from two random galaxies in a given environment. Considering the observed fraction of galaxies in each LSS environment that are late-type, we can define $P_{\rm{LSS}}(\rm{L})$ as the probability of a galaxy in a given environment being late-type. The same applies to early-types, that would be merely $P_{\mathrm{LSS}}(\mathrm{E})=1 - P_{\mathrm{LSS}}(\mathrm{L})$. With those defined probabilities, and assuming that these probabilities are independent on being the most massive or least massive galaxy in the pair, we can define the expected combinations in each LSS environment (i.e., $P_{\mathrm{LSS}}(\mathrm{EL})=P_{\mathrm{LSS}} (\mathrm{E}) \cdot P_{\mathrm{LSS}}(\mathrm{L})$). These baseline probabilities appear in the upper row of pie charts in Figure \ref{fig:piecharts}, while the upper panel histograms represent the T-Type distributions per LSS, which are used to extract the aforementioned probabilities. In the last row appear the observed combinations. These are obtained through the real, observed pairs. The observed fractions agree closely with the random model when inspecting same-type pairs (LL and EE): the largest discrepancy involves a 4\% difference between EE pairs in voids. The different-type pairs cannot be compared directly; however, when considering the total fraction of them (EL+LE) the distributions are congruent with one another. Considering only the different-type pairs, the observed distributions in the two LSS environments yield a higher proportion of EL pairs than LE pairs -- in particular, while the random sampling assumes a 50/50 chance of the central galaxy being early- or late-type, we find that in both voids and NCNV $\sim$70\% of centrals in pairs are early-type ($70.4\%$ and $71.1\%$, respectively). This shows that early-type galaxies tend to be the central galaxy in a pair, independently of the LSS environment.

\section{Discussion}\label{sec:discussion}

\subsection{Fractions of singlets and multiplets in the LSS}

As seen in Table \ref{tab:percentages}, singlets are heavily associated to low density environments: in voids, they make up the $\sim$80\% of the galaxies, a percentage 20\% higher of what is found in NCNV. Galaxies with more than three companions tend to inhabit high-density environments, making up 27\% of the total sample in NCNV, and being the main contributors in clusters with 70\% of the galaxies. The latter is to be expected, as it reflects the underlying distribution of galaxies within clusters. In the case of small groups, both pairs and triplets tend to inhabit lower density environments. Particularly, pairs are clearly inclined towards void environments: they are more than a third of all multiplets in voids (38\%). In NCNV, they make up a fourth of the total of multiplets (24\%). Triplets, however, do not show this preference towards voids, but slightly more so towards NCNV: 9\% of multiplets in NCNV are in triplets, in contrast to the 4\% in voids. These results are in agreement with \cite{2020MNRAS.493.1818D}, that suggested that triplets favour denser LSS environments than pairs, but less dense environment than larger groups; their sample, however, comprised compact groups. We find the same result for systems that are not necessarily compact, nor isolated. In clusters, as it is expected, $90\%$ of galaxies have more than one or two neighbours. The presented fractions of multiplets are prone to be underestimated due to the fibre collision problem in SDSS data, as well as the completeness bias and the consequent loss of low-mass galaxies in this sample, as discussed in Sect. \ref{sec:completeness}. 

We now compare these results to the findings of previous works in galaxy environment. In \cite{2007ApJ...671..153Y}, they employ a halo-based group finder to identify groups in SDSS DR4 in the redshift interval ${0.01<z<0.20}$. Making use of their main sample (denoted sample II) they find that 73.6\% of galaxies are singlets; a value only comparable to our fraction of singlets in voids, but far from the percentage of singlets that we find in our sample (59.4\%). Furthermore, the discrepancy between the two values would be increased by a margin of $\gtrsim 10\%$ if we were to consider companions of $\log M_\star / \mathrm{M}_\odot < 9.5$, as we would find a lower fraction of singlets. In \cite{2022A&A...668A..69E}, they employ a global density-luminosity field to characterise large-scale environment, classify local environment using \cite{2012A&A...540A.106T, 2014A&A...566A...1T}, and then provide fractions in different environments in Table 1 for galaxies in the redshift range $0.009<z<0.200$, and within a volume-limited selection. A comparison between the two studies is challenging due to the inherent differences in the samples' environment classification, and redshift range. However, both studies utilise a volume-limited sample to derive statistical results, thereby ensuring a degree of comparability -- furthermore, we believe it is valuable to compare results obtained through different classification methods. When inspecting the fraction of singlets they find very compatible results with this work: 73\% in the least dense environments ($D8<1$, analogous to voids), and 17\% in the most ($D8>5$, analogous to our cluster galaxies definition). They additionally provide fractions of what they call very old (VO) galaxies, which correspond to galaxies with $D_n(\rm{4000)}\geq1.75$ and contain mainly old stellar populations; it is, however, not straight-forward to compare VO galaxies to our early-type classification, as $D_n(\rm{4000)}$ relies on spectral information and T-Type is purely photometric. Although most of the galaxies do not fall within these categories, using $D_n(\rm{4000)}$ excludes low-mass early-type galaxies and includes massive late-type galaxies in the sample. Even within these limitations, the results do not differ more than a 15\%: where we find 50\% of early-type singlets in clusters, they find 36\%. In voids, these numbers are 30\% and 27\%, respectively. The compatibility of both studies, despite the differences in method, emphasises how robust are the differences of the galaxy populations between large-scale environments. The systematic finding of a higher proportion of early-type galaxies and multiplets may be a consequence of our consideration of morphological completeness (Appendix \ref{appendix:et-bias}), given that low-mass early-type galaxies are prone to being overlooked in seemingly complete samples.

\subsection{Singlets and multiplets}\label{sec:discussion-local}

The stellar mass distribution in the different LSS environments reflects what was noted in various earlier studies: galaxies in voids tend to, overall, lower masses \citep{2004ApJ...617...50R, 2005ApJ...624..571R}. We find that singlets in voids and NCNV tend to be less massive than galaxies in groups in those environments. In clusters, the mass distributions are similar for singlets and multiplets. It is not clear which mechanisms could cause those mass differences (or lack thereof), but either way, it would be hasty to think of them as independent from the LSS. As seen in Figure \ref{fig:effr_frac}, singlets in clusters are found in the outskirts, in contrast to multiplets. This finding could be interpreted as a chronology for cluster galaxies, and indicates that most of the infalling singlets in clusters will have nearby companions during their stay in such an overdense environment. Our LSS characterisation, however, does not consider complex kinematical situations such as fly-by or backlash galaxies, that are definitely present in clusters. Clusters are extreme environments in which galaxies reach high velocities and can suffer from severe harassment \citep{1996Natur.379..613M} and other intracluster phenomena, whether their sources are tidal events or hydrodynamical physics. Simulations suggest that these phenomena lead to stellar mass loss of up to the 50\% in the cluster centre \citep{2015A&A...576A.103B}; we do not find evidence that the stellar mass distribution of singlets (located in the outer regions) and multiplets (in the inner parts) are different. Therefore, we do not find particular evidence of stellar mass loss. As mentioned, a more extensive study would be required to solve this question in particular, along with a deeper sample that would allow the study of dwarf galaxies.

Other simulations \citep{2005MNRAS.364..607M, 2012MNRAS.424.2401V} also predict a morphological transformation of discs entering cluster environments, although this transformation is highly dependent on orbital parameters, such as inclination of the disc with respect to the orbital plane. We find that multiplets are a 17\% more frequently early-type than singlets, being the latter in the outskirts. This is consistent with galaxies transforming their morphology when entering the cluster. Furthermore, this scenario is congruent with the finding of late-type galaxies in clusters showing smaller sizes and higher concentrations than those in voids and NCNV \citep{2025A&A...695A..84P}. Although further evidence would be required to confirm this relocation scenario, it is relevant that 50\% of singlets in clusters, despite inhabiting mostly the outskirts, show early-type morphologies -- nearly the same fraction of early-types as found in NCNV multiplets (44\%). This, and the nearly identical stellar mass distribution of singlets in NCNV and clusters, supports the idea of singlets in clusters being galaxies in transition from a nearby filament, and these galaxies are likely subject to preprocessing. This preprocessing could be related to the filament environment itself \citep{2018MNRAS.474..547K, 2024MNRAS.529.2595B}, where local densities are possibly the main driver of quenching \citep{2022A&A...657A...9C}.

In the case of voids, the effect of the group environment is of particular interest, as the mechanisms that assemble groups together in such underdense areas of the Universe are not yet understood. Our results suggest that void galaxies in different local densities not only have similar stellar mass distributions (although statistically distinct), but are distributed in a similar way with respect to the void centres, too (Figure \ref{fig:effr_frac}), in agreement with \cite{ArgudoFdez_submitted}, where galaxy groups are characterised in voids. These two facts suggest that both singlets and groups may have formed similarly in voids. This could mean that there is no systematic mechanism that relocates groups from NCNV to voids. Our results are in agreement with \cite{2024A&A...692A.258A}, that studied morphologies in void galaxies to find that they are independent on the local number densities, the size of the void, and their position with respect to the centre of the void -- in opposition to filaments, where local densities have been discussed to be main evolutionary drivers \citep{2022A&A...657A...9C, 2024MNRAS.534.1682O}.

In summary, we find that both local and large-scale environment changes modify slightly both stellar mass and morphology distributions. These effects are likely not always separable, as density is a crucial part of the definition of a LSS; but they appear to be to some extent, as removing the local effects does not eliminate the large-scale effects on the distributions, and vice versa. It is significant that morphology, being a fundamental tracer of environmental density \citep{1980ApJ...236..351D}, is dependent, too, on the LSS: it may be taken as an indicator that the morphology-density relation can be extended towards larger scales. It is, however, out of the scope of this work to disclose the precise relation of the large-scale environment density with morphology. It should be taken into consideration that morphology in this work is studied in samples that are not matched in stellar mass.

\subsection{Galaxy pairs}

% stellar mass

As seen in Figure \ref{fig:mn_mc}, we find that pairs in voids show a tendency towards lower $M_n/M_c$ fractions than pairs in NCNV. Since the stellar mass distribution of the central galaxies is fixed, this difference is exclusively due to satellites in voids being less massive than in denser environments. Still, we also find central galaxies to be less massive in voids than in denser environments. We will discuss this further in Section \ref{sec:halos}.

% morphology segregation

In terms of morphology, central galaxies in pairs tend to be more early-type, independently of LSS, and in contrast with our simple random model (Figure \ref{fig:piecharts}). This is to be expected, as the main assumption when drawing the random distributions was that morphology is independent of the galaxy being central or satellite. We find that assumption to be wrong, both in our results (see Figure \ref{fig:ttype-pairs}) and in many previous studies that find early-type galaxies to be, overall, more massive than late-type galaxies. Therefore, simply by virtue of the definition of the central galaxy as the most massive in the pair, there are more EL than LE pairs in every environment, since the most massive galaxy in a pair of an early-type and a late-type galaxy is most usually the early-type galaxy. Consequently, we can only compare the total fraction of different-morphology pairs (EL+LE) when contrasting the random and observed distributions.

In pairs of the same morphological type, there is a high level of agreement between observed pairs and the random distributions. This suggests that the formation of pairs with different morphologies essentially depends on the availability of galaxies in a given LSS environment, and that galaxies forming pairs do not necessarily have undergone any kind of preprocessing that could modify their morphology beforehand. However, we expect this result to be dependent on our local environment characterisation: we prioritised criteria that assured physical bounding between galaxies \citep{2014A&A...564A..94A}, although according to previous works, galaxies in pairs require closer sky-plane distances to modify each other's properties (i.e., star formation rate or morphology) significantly \citep[among others, ][]{2010MNRAS.407.1514E, 2012MNRAS.426..549S, 2013MNRAS.433L..59P}. Additionally, in this study it was necessary to exclude all the dwarf regime to assure completeness, being the low-mass galaxies the more prone to be subject to morphological transformations. This kind of study would benefit from the availability of deeper wide-field galaxy surveys, as dwarfs are highly useful laboratories to study environmental influence. Nevertheless, we found in \cite{2024A&A...691A.341T} that, under our local environment classification, there is evidence of a slight delay ($\sim 1 \, \rm{Gyr}$) in the star formation histories of singlets with respect to multiplets, despite the limitations in stellar mass. We suggest that both this work and \cite{2024A&A...691A.341T} point towards a common evolutionary origin of multiplets, rather than physical mechanisms modifying the morphology of galaxies at such separations.

\subsection{Halos and the LSS}\label{sec:halos}

There is a fundamental difference when it comes to void galaxies, as they are systematically less massive than galaxies in other environments, independently of their local environment (Figure \ref{fig:pp-mass}). The physical reasons that cause this shift in mass are still in study, but as it appears to be a large-scale phenomenon, it is reasonable to point towards halo formation in different LSS contexts.

Under the current cosmological paradigm, the skeleton of the cosmic web, as we know it in the present, is the end product of density perturbations in the early Universe. As predicted by simulations, void (cluster) environments are regions in space that appear where density waves of medium and large-scales combine in similar under-density (over-density) phases \citep{2011A&A...531A..75E}. Those areas are populated by dark matter halos, where later galaxies assemble. It was soon noticed that the clustering in halos (at fixed halo mass) is not independent of the LSS \citep{2005MNRAS.363L..66G}, as older halos cluster more strongly. This is referred as the halo assembly bias. Studies in the last decade have put the focus on the relation between halos and the LSS, as well as their relation to galaxy evolution \citep{2017MNRAS.470.3720T, 2018ApJ...853...84Z, 2018MNRAS.480.3978A, 2020A&A...638A..60A}. These studies have found that, in voids, halos generally are younger (more recently formed) than in denser environments. Halos in voids are less likely to host central galaxies, and show a lower occupation of satellites, which is consistent with our results (Table \ref{tab:percentages}).

Many works put importance in the power of the halo occupation distribution (HOD) as a tool to describe galaxies in the context of their dark matter halo. The HOD describes the probability that a virialised halo of certain mass contains $N$ galaxies, and since this framework naturally associates the halo population with halo mass, the existence of the halo assembly bias motivates the idea that the halo assembly history may affect the properties of the galaxies within -- this is referred as the galaxy assembly bias. Within this context, \cite{2020A&A...638A..60A} studies the HOD in voids in simulations and finds that, in voids, central galaxies are $\sim$10\% less massive than in other environments, and the average mass per satellite is $\sim$30\% lower. This is consistent with our results regarding galaxy pairs (Figure \ref{fig:mn_mc}), as we find that centrals and satellites in voids have higher differences in stellar mass, but also explains the shift towards lower values in stellar mass distributions of void galaxies.

Many results, both in simulations and observations, point towards the relevance of the LSS, on its own, in the evolution of galaxies. As meaningful examples, \cite{2025A&A...695A..84P} found the mass-size relation of galaxies to be affected by the LSS. Low-mass late-type galaxies are less massive in clusters than in NCNV and voids, while early-types are less massive in voids than in denser environments. \cite{2023Natur.619..269D} found that galaxies in voids have a slower stellar mass assembly when inspecting their star-formation histories, and this result stands independently of local environment \citep{2024A&A...691A.341T}. This difference in assembly rate has also been suggested by simulations, as it appears that galaxies in voids experience roughly the same number of mergers as denser environments, although later in time \citep{2024MNRAS.528.2822R}, with more of those mergers being minor \citep{2022MNRAS.517..712R}. The higher presence of minor mergers in voids is also consistent with our finding of lower neighbour-to-central mass ratio in void pairs (Figure \ref{fig:mn_mc}), as well as with groups in voids being in an early stage of their evolution \citep{ArgudoFdez_submitted}. Ultimately, this work contributes to all these previous studies by providing more observational evidence that the LSS is a meaningful agent in the evolution of galaxies. We additionally suggest that a framework that contemplates the halo and galaxy assembly biases is needed to reach a comprehensive view of galaxy formation and evolution along the cosmic web.

\section{Summary and conclusions}\label{sec:conclusions}

In this study, we classified a sample of $25\,309$ galaxies with $\log M_\star/\mathrm{M}_\odot > 9.5$ in $0.02 < z \leq 0.04$ according to their large-scale and local environments. This range in redshift and limitation to high-mass galaxies assures completeness not only in mass, but also in morphology for this sample based on SDSS data. Galaxies were classified in terms of the main large-scale structures (LSS); as void galaxies if they appear in \cite{2012MNRAS.421..926P} void catalogue, in clusters if they are part of groups of more than 30 galaxies in \cite{2017A&A...602A.100T}, and as being part of intermediate-density environments if they are not part of either clusters nor voids (NCNV). They are classified in terms of local environment as well, as singlets or multiplets, if there is a galaxy within velocity differences $\Delta v \leq 160 \ \mathrm{km \ s}^{-1}$ and sky-projected distances $\Delta r_p \leq 450 \ \mathrm{kpc}$ (being these conditions for physical bounding). We explored the distributions of stellar mass and morphology of singlets and multiplets in different LSS, studying the case of pairs in particular for voids and NCNV, and differentiating between central and satellite galaxies. The main results can be summarised as follows:

\begin{enumerate}
    \item The position of singlets and multiplets within the LSS is compiled in Table \ref{tab:percentages}. In voids, $\sim22\%$ of galaxies inhabit pairs, triplets, or groups. This fraction grows for NCNV and cluster galaxies. The denser the large-scale environment, the greater the proportion of higher-order multiplets: galaxies in pairs make up more than a third of all the multiplets in voids, a fourth of those in NCNV, but only around a fifteenth of those in clusters. Galaxies in triplets, however, favour slightly more NCNV than voids.

    \item The denser the environment, the greater the fraction of early-type galaxies. This applies to the local environment, as anticipated by the morphology-density relation \citep{1980ApJ...236..351D}, but also to the large-scale environment. In voids, 30\% of singlets are early-type, in opposition to the 34\% in NCNV and 50\% in clusters. The same thing occurs with late-type galaxies: a higher fraction of multiplets are late-type galaxies in voids than in NCNV and clusters. These fractions are comprised in Table \ref{tab:ttype-local}. This result suggests that the morphology-density relation could be expanded towards the large-scale structures, as the local densities are not fully responsible of the morphology segregation in the cosmic web.
    
    \item Void galaxies are notably less massive than galaxies in NCNV and clusters, independently of the local environment. Galaxies in NCNV and clusters show similar stellar mass distributions, with the exception of multiplets in clusters having a meaningful contribution of less massive galaxies.

    \item Fixing the stellar mass distributions of central galaxies in pairs, we find that pairs in voids exhibit lower neighbour-to-central mass ratio, i.e., we find a larger difference in mass between central and satellites. Both centrals and satellites in voids are less massive than in denser environments.

    \item The morphological segregation in pairs appears to be solely related to the availability of galaxies with different morphologies in each LSS environment. Despite central galaxies tending to be early-type, observations are in agreement with a simple random sampling scenario, in which two random galaxies in a certain LSS environment are put together in a pair.
\end{enumerate}

The main bias of this work when presenting representative results for the local Universe is the exclusion of low-mass galaxies ($\log M_\star/\mathrm{M}_\odot < 9.5$). It is likely that dwarf galaxies play a meaningful role when studying the LSS \citep[see][as a meaningful example]{2025A&A...698A.260B}, accounting for the tendency of voids towards the low masses, and the inherent differences in evolutionary rates in the cosmic web \citep{2023Natur.619..269D}. However, as discussed in Appendix \ref{appendix:et-bias}, statistical studies in extragalactic astrophysics are naturally limited by the lower brightness of early-type galaxies at a given stellar mass, which threatens completeness in morphology. The advent of large, deeper galaxy surveys is crucial to bridge this distance and accurately portray the local Universe. In this work, finding slightly larger fractions of early-type galaxies in all environments than previous studies is a good indicator that our sample is complete in the studied range of stellar masses and provides new insight on the galaxy populations that inhabit our galactic surroundings.

Regarding the large-scales, the results presented in this work are consistent with the LSS environment having an undeniable effect on the formation and evolution of galaxies in the Universe. While the local environment is responsible for some observed properties, especially in close pairs, compact groups, or mergers (which are not addressed in this work), observations such as stellar mass distributions in voids cannot be explained by local density alone. In the case of void galaxies in particular, these results seem to be related not to what the galaxies experience during their lifetime, but to the conditions under which they were born. This suggests that, in void galaxies, nature plays a greater role than nurture in the nature-versus-nurture debate. These galaxies appear to share a common evolutionary origin, possibly relating to their assembly within dark matter halos in the early Universe. Further clarification on this topic can be found by targeting void galaxies as a case study, with initiatives such as the CAVITY project \citep{2024A&A...689A.213P, 2024A&A...691A.161G}.

% \section{Data availability}
% Tables xx and xx are only available in electronic form at the CDS via anonymous ftp to \url{cdsarc.u-strasbg.fr} (130.79.128.5) or via \url{http://cdsweb.u-strasbg.fr/cgi-bin/qcat?J/A+A/}.

\begin{acknowledgements}
We are grateful to the anonymous referee, whose suggestions helped the authors to improve this manuscript. G.T.R. acknowledges financial support from the research project PRE2021-098736 funded by MCIN/AEI/10.13039/501100011033 and FSE+. S.V., I.P. and B.B. acknowledge financial support from
the grant AST22.4.4, funded by Consejería de Universidad, Investigación e
Innovación and Gobierno de España and Unión Europea – NextGenerationEU,
and by PID2023-149578NB-I00, financed by MCIN/AEI. M.A-F. acknowledges support from ANID FONDECYT iniciación project 11200107 and the Emergia program (EMERGIA20\_38888) from Consejería de Universidad, Investigación e Innovación de la Junta de Andalucía. S.D.P. acknowledges financial support by MINECO under grants PID2023-149578NB-100 and PID2022-136598NB-C32. Y.G.K. acknowledges financial support from PREP2023-001684 funded by MCIU/AEI/10.13039/501100011033 and the FSE+.

This work made use of Astropy:\footnote{\url{http://www.astropy.org}} a community-developed core Python package and an ecosystem of tools and resources for astronomy \citep{astropy:2013, astropy:2018, astropy:2022}; JupyterLab\footnote{\url{https://github.com/jupyterlab/jupyterlab}}; matplotlib \citep{Hunter:2007}; SciPy, a collection of open-source software for scientific computing in Python \citep{2020SciPy-NMeth}; and NumPy, a structure for efficient numerical computation \citep{harris2020array}. 

Funding for the SDSS and SDSS-II has been provided by the Alfred P. Sloan Foundation, the Participating Institutions, the National Science Foundation, the U.S. Department of Energy, the National Aeronautics and Space Administration, the Japanese Monbukagakusho, the Max Planck Society, and the Higher Education Funding Council for England\footnote{\url{http://www.sdss.org/}}. 

Funding for SDSS-III has been provided by the Alfred P. Sloan Foundation, the Participating Institutions, the National Science Foundation, and the U.S. Department of Energy Office of Science\footnote{\url{http://www.sdss3.org/}}. SDSS-III is managed by the Astrophysical Research Consortium for the Participating Institutions of the SDSS-III Collaboration including the University of Arizona, the Brazilian Participation Group, Brookhaven National Laboratory, Carnegie Mellon University, University of Florida, the French Participation Group, the German Participation Group, Harvard University, the Instituto de Astrofisica de Canarias, the Michigan State/Notre Dame/JINA Participation Group, Johns Hopkins University, Lawrence Berkeley National Laboratory, Max Planck Institute for Astrophysics, Max Planck Institute for Extraterrestrial Physics, New Mexico State University, New York University, Ohio State University, Pennsylvania State University, University of Portsmouth, Princeton University, the Spanish Participation Group, University of Tokyo, University of Utah, Vanderbilt University, University of Virginia, University of Washington, and Yale University.
\end{acknowledgements}

\bibliographystyle{aa} % style aa.bst
\bibliography{paper} % your references Yourfile.bib

\begin{appendix}
% \onecolumn
\section{The dependence of mass completeness on galaxy morphology}\label{appendix:et-bias}

The present Appendix is dedicated to the brief description and parametrisation of completeness, with the objective of accounting for both stellar mass and morphology biases. In this Section we will make use of the parent sample of this study, based on the main galaxy sample of the SDSS DR12 -- nonetheless, this analysis is of crucial importance for any work in extragalactic statistics, especially when working with wide-area surveys, that are relatively shallow.

In any magnitude-limited galactic survey, the galaxy detections are biased towards the brighter end of the real galaxy population. This naturally implies that the less massive (fainter) galaxies will not be accounted for in a random survey. Furthermore, this effect will be more serious the further we intend to explore, as luminosity decreases quadratically with distance. In this context, completeness has been heavily studied and taken into account to provide science-sufficient observations in many studies and collaborations of many (if not all) astrophysical fields. 

However, a more nuanced explanation is required to assure real completeness, as morphology (or a convenient tracer, such as colour), is related to brightness in galaxies. For a fixed mass, the more early-type are the galaxies, the fainter they are, as the absence of young stellar populations takes an evident toll on the total brightness of a galaxy. Therefore, the fraction of missed galaxies not only depends on redshift and stellar mass, but also on morphology. This dependence complicates the task of portraying the galaxy populations in the Universe, as our sampling of these populations is naturally biased towards an excess of blue objects, in the same way it is naturally biased towards more massive galaxies.

In Figure \ref{fig:completeness}, we portray completeness curves while binning the parent sample of this study in both T-Type and colour. It is straightforward to see that the completeness of the samples is strongly affected by morphology. As a meaningful example, if we intend to study very blue, star forming galaxies at redshift $z\sim0.05$, a stellar mass cut in $\log M_\star / \mathrm{M}_\odot \sim 9.0$ is a reasonable choice: however, if quenched, elliptical galaxies are object of interest, a complete sample would entail considering $\log M_\star / \mathrm{M}_\odot > 9.75$, definitely missing all the dwarf galaxy population. Dwarf galaxies are the most affected by this bias; however, disregarding them affects any statistical study in extragalactic astrophysics, as it is this work. 

Note that this issue has been previously reported. \cite{2021MNRAS.500.4469T} stressed that not accounting for the fractions of quenched, green valley, or star-forming per environment could introduce severe selection biases and maximise the apparent influence of environment. \cite{2025MNRAS.538..153K}, who studied fractions of red dwarfs in the local Universe, emphasised the need for deep surveys even in low redshift regimes for this same reason.

\begin{figure*}
    \centering
    \begin{subfigure}[t]{0.49\textwidth}
    \includegraphics[width=1\linewidth]{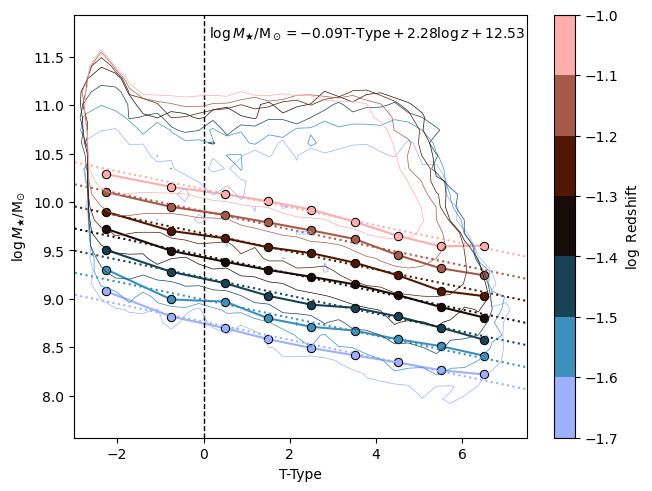}
    \caption{Data binned by T-Type, provided by \cite{2018MNRAS.476.3661D}. The vertical dashed line marks the separation between early- and late-type galaxies.}  
    \label{fig:completeness_fit_TType}
    \end{subfigure}
    \begin{subfigure}[t]{0.49\textwidth}
    \includegraphics[width=1\linewidth]{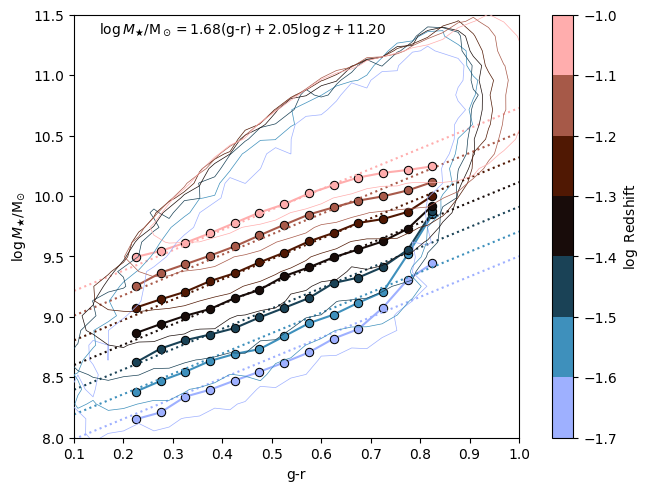}
    \caption{Data binned by $g-r$ colour, provided by the SDSS DR12.}
    \label{fig:completeness_fit_colour}
    \end{subfigure}
\caption{Parametrisation of completeness. Both panels show stellar mass as a function of morphology for the parent sample. Galaxies are binned in $\log z$ as shown by the colourbar, with $0.02 \leq z \leq 0.10$. For each $\log z$ bin, each line corresponds to completeness (computed as the 5\textsuperscript{th} percentile of the stellar mass in the shown bins of T-Type/colour). Dotted lines correspond to the linear fit portrayed at the top of each panel. The background distributions represent the 5\% contour of the galaxy distribution at each $\log z$ bin.}
\label{fig:completeness_fit}
\end{figure*}

We are able to parametrise the relation between stellar mass, T-Type (or colour), and redshift in a complete sample. The minimum stellar mass cut required to obtain a complete sample has a linear dependence with T-Type (or colour) and the decimal logarithm of redshift ($\log z$). This is shown in Figure \ref{fig:completeness_fit}. 

To obtain the parametrisations, we fit completeness (Figure \ref{fig:completeness_fit}) to a linear $\log M_{\star,\text{min}} = a \times \mathcal{M} + b\log z + c$ model, where $a$, $b$, and $c$ are parameters, $\mathcal{M}$ denotes the morphology proxy (T-Type or $g-r$ colour), and $\log M_{\star,\text{min}}$ is the lower mass limit required for the sample to be complete. To obtain a truthful fit, we fit recursively using different $\mathcal{M}$ bins in each iteration. From the ones with successful fits, we exclude the outliers, defining a cut of 2.5 on the distance to the median of the fit parameters (where the distance is normalised by the median value of the parameter). Then, the final parameters are the median of the successful, trustworthy values (19 in the case of the T-Type fitting, 55 for the colour fitting). The final parametrisations are shown in Table \ref{tab:params}.

When we apply this parametrisation to ${\text{T-Type} = -2.36}$ (the 5\textsuperscript{th} percentile of the T-Type distribution in our sample) and ${z=0.04}$, we obtain a stellar mass low-limit of ${\log M_\star / \mathrm{M}_\odot = 9.56}$. Subsequently, we apply a ${\log M_\star / \mathrm{M}_\odot = 9.5}$ cut and assure the completeness in this study (tests were performed with the cut at ${\log M_\star / \mathrm{M}_\odot = 9.6}$, with no meaningful variations).

\begin{table}[]
\caption{Fit parameters of completeness parametrisations}
    \centering
    \begin{tabular}{lccc}
        \hline \hline
        Morphology & $a$ & $b$ & $c$ \\
        parameter ($\mathcal{M}$) &&&\\
        \hline
        T-Type & $-0.093 \pm 0.001$ & $2.28 \pm 0.02$ & $12.53 \pm 0.03$\\
        $g-r$ & $1.68 \pm 0.02$ & $2.05 \pm 0.01$ & $11.20 \pm 0.03$ \\ \hline
    \end{tabular}
    \tablefoot{In these fittings, we assume the decimal logarithm of redshift ($\log z$) and the morphology proxy (T-Type or $g-r$) as independent variables as follows: $\log M_{\star,\text{min}} = a \times \mathcal{M} + b\log z + c$, where $\mathcal{M}$ is the morphological tracer, $M_\star$ units are $\mathrm{M_\odot}$, and $a$, $b$, and $c$ are parameters. $\log M_{\star,\text{min}}$ denotes the minimum mass cut required for the sample to be complete in both mass and morphology (i.e., the 95\% of the sample is included). To appropriately apply these parametrisations, this mass cut should be calculated making use of the higher redshift of the sample, and the lower (higher) T-Type ($g-r$ colour). Since these parameters were obtained fitting iteratively over several bins on the morphological tracer, the errors in this table are the standard deviations of the parameters in all the successful fittings.}
    \label{tab:params}
\end{table}

We briefly justify the form of the parametrisation by considering the approximation ${\log M_\star/L_{b} \propto \mathcal{M}}$, being $L_b$ the luminosity of a galaxy in a certain band $b$. Parting from the relation between absolute magnitude in $M_b$ and distance to the galaxy $d$:

\begin{equation}
\begin{split}
    M_b & = m_b - 5\log d \, [\rm{pc}] + 5 \\
    & = m_b - 5\log z - 5\log\left(\frac{c}{H_0} \, [\rm{pc}]\right) + 5 \quad ,
\end{split}
\label{eq:M-z}
\end{equation}

\noindent where $m_b$ is the apparent magnitude in the same given band, and $c$ is the speed of light. At $z>0.02$, we are able to approximate $d$ through the Hubble law and the usual definition of redshift (${z=v/c}$, being $v$ the recession velocity of the galaxy). We consider, too, the relation between absolute magnitude and luminosity:

\begin{equation}
    M_b-M_{b,\odot} = -2.5\log \frac{L_b}{\mathrm{L_\odot}} \quad ,
\label{eq:M-L}
\end{equation}

\noindent where $M_{b,\odot}$ is the solar absolute magnitude in the $b$ band. Joining both Equations \ref{eq:M-z} and \ref{eq:M-L}, arises the relation between luminosity and redshift:

\begin{equation}
    \log \frac{L_b}{\mathrm{L_\odot}} = 2\log z - 0.4m_b + C \quad ,
\label{eq:L-z}
\end{equation}

\noindent with $C = 2\log\left(\frac{c}{H_0} \, [\rm{pc}]\right) - 2 + 0.4M_{b,\odot}$ being a constant. If $\log M_\star/L_b = f(\mathcal{M})$, we can then state:

\begin{equation}
    \log\frac{M_\star}{\mathrm{M}_\odot} = f(\mathcal{M}) + 2\log z - 0.4m_b + C \quad .
\end{equation}

The mass-luminosity relation $\log M_\star/L_b$ in known to be approximately linear with colour \citep[for example,][]{2003MNRAS.341...33K}, although it does not express that clear behaviour with T-Type. However, we can state that the behaviour is linear when considering completeness with $\mathcal{M}$ (i.e., the 5\textsuperscript{th} percentile of the stellar mass per bins of T-Type/colour), as shown previously in Figure \ref{fig:completeness_fit}. This and the aforementioned analysis justifies parametrising completeness as $\log M_{\star,\text{min}} = a \times \mathcal{M} + b\log z + c$.

% -------------------------------------

\section{Separation and velocity differences in pairs}\label{appendix:pairs}

\begin{figure*}
    \centering
    \includegraphics[width=\textwidth]{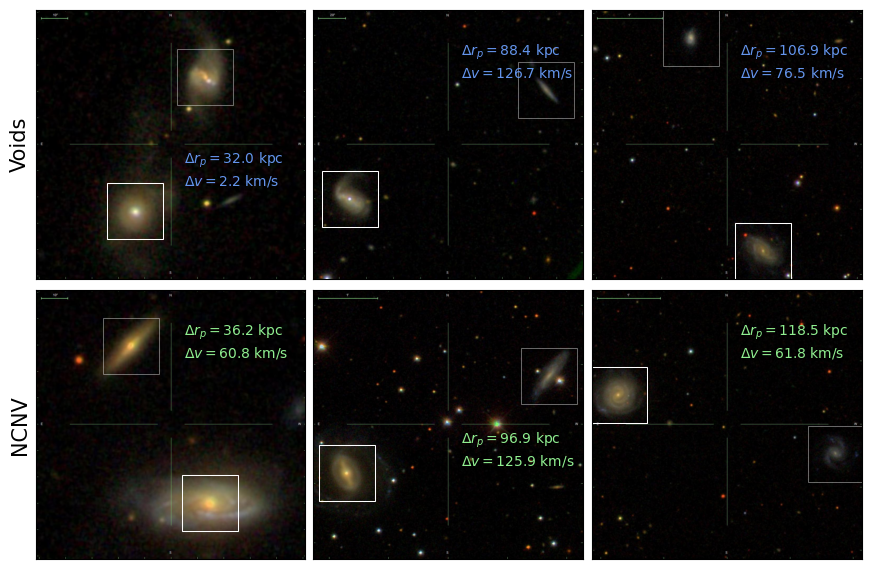}
    \caption{Selected pairs from voids (upper row) and NCNV (lower row). The galaxies in the pairs are denoted by white squares of the same size, being the brighter square the central (more massive) galaxy of the pair. \textit{From left to right}, the columns portray pairs with on-sky separations $\Delta r_p < 50 \, \rm{kpc}$, $\Delta r_p < 100 \, \rm{kpc}$ and $\Delta r_p > 100 \, \rm{kpc}$, respectively. All panels are three-band (\textit{gri}) images extracted from the SDSS DR18 SkyServer\protect\footnotemark. The \texttt{GalID}s \citep{2017A&A...602A.100T} of the central galaxy in each pair are the following, from top to bottom: 296698, 191424 and 358261 in voids, and 31694, 118572 and 113219 in NCNV.}
    \label{fig:examples-pairs}
\end{figure*}
\footnotetext{We made use of the SciServer Python package. More information available at \url{https://www.sciserver.org/docs/sciscript-python/SciServer.html}.}

\begin{figure}
\centering
\includegraphics[width=\linewidth]{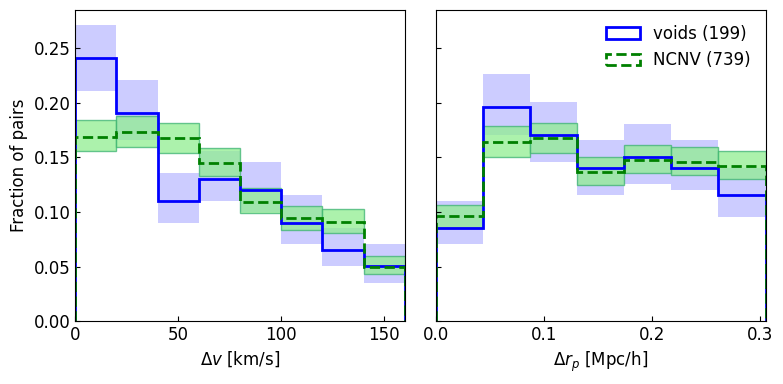}
\caption{Distributions of velocity differences \textit{(left panel)} and sky-projected distances \textit{(right panel)} in pairs in the different LSS environments (voids in blue solid lines and NCNV in green dashed lines). The numbers in the legend correspond to the number of pairs per environment. The shadowed areas correspond to $1\sigma$ bootstrapping uncertainties.}
\label{fig:tempel-pp-dv-pd}
\end{figure}

In this Appendix, we briefly show the velocity differences ($\Delta v$) and on-sky separations ($\Delta r_p$) of the void and NCNV pairs in the sample and discuss their effect on the local environment classification. In Figure \ref{fig:examples-pairs} appear examples of galaxy pairs.

Figure \ref{fig:tempel-pp-dv-pd} shows the distributions of $\Delta v$ and $\Delta r_p$ in void and NCNV pairs. In voids, galaxies in pairs exhibit lower velocity differences with their satellite than galaxies in NCNV. In sky-projected distances, the distribution of NCNV pairs appears to be slightly flatter, too, than the one in voids -- although it is more difficult to state whether this is a physical effect, as in most of the bins the differences lay within the range of uncertainties.

Dynamical properties of pairs in voids differ from pairs in NCNV, as pairs in NCNV show larger velocity differences. Although, as discussed, we exclude cluster pairs from this study, it is expected that the effect of the potential well in these extreme environments will modify the dynamical behaviour of galaxies in groups even more so. These results are consistent with previous studies in different kind of galaxy groups, such as compact groups and associations of dwarf galaxies, that all find larger velocity dispersions with large-scale environmental density \citep{2023MNRAS.520.6367T, 2023MNRAS.525..415Y}. 

This trend arguably has an impact on the way physical bounding of groups should be constricted in the different LSS environments. In \cite{2015A&A...578A.110A}, evidence is found of physical bounding of pairs and triplets in isolation under velocity differences $\Delta v \leq 160 \, \mathrm{km \cdot s^{-1}}$ and projected distances $\Delta r_p \leq 0.45$ Mpc, which are the criteria we apply in this work. However, the tendency towards higher velocity differences in pairs inhabiting high density environments, and non necessarily isolated systems, suggests that these criteria are inherently conservative in these environments. Although extensive work has been carried out on galactic local environment by selecting the samples through criteria of projected separation and line-of-sight velocity differences, hardly ever have these criteria been considered out of the domain of close pairs and merger systems \citep[][among many others]{2000ApJ...536..153P, 2008AJ....135.1877E, 2019A&A...631A..87V}. Accommodating these criteria to the needs of each LSS becomes crucial in the task of studying the extent of the influence of local and global environments on galaxy evolution.

\end{appendix}

\end{document}